\documentclass[12pt]{article} 
\usepackage[hidelinks]{hyperref}
\usepackage{graphicx,multirow}
\usepackage{amssymb,amsmath,amsfonts,bm}
\usepackage{setspace}
\usepackage{natbib}
\usepackage{multicol}
\usepackage{tabularx}
\usepackage{array}
\usepackage{caption}
\usepackage{xcolor}
\usepackage[misc]{ifsym}
\usepackage{colortbl}
\usepackage{titlesec}
\usepackage{enumitem}

\newcommand{\be}{\begin{equation*}}
\newcommand{\ee}{\end{equation*}}
\newcommand{\bne}{\begin{equation}}
\newcommand{\ene}{\end{equation}}

\begin{document}


\title{A General Simulation-Based Optimisation Framework for Multipoint Constant-Stress Accelerated Life Tests}
\author{Owen McGrath\footnote{Corresponding author. University of Limerick, Ireland; owen.mcgrath@ul.ie}  \hspace{3cm}
Kevin Burke\footnote{University of Limerick, Ireland; kevin.burke@ul.ie \newline\newline\indent\indent
The corresponding author acknowledges the financial support of the Technological University of the Shannon.}}
\date{\today}

\maketitle

\begin{abstract}

Accelerated life testing (ALT) is a method of reducing the lifetime of components through exposure to extreme stress. This method of obtaining lifetime information involves the design of a testing experiment, i.e., an accelerated test plan. In this work, we adopt a simulation-based approach to obtaining optimal test plans for constant-stress accelerated life tests with multiple design points. Within this simulation framework we can easily assess a variety of test plans by modifying the number of test stresses (and their levels) and evaluating the allocation of test units.  We obtain optimal test plans by utilising the differential evolution (DE) optimisation algorithm, where the inputs to the objective function are the test plan parameters, and the output is the RMSE (root mean squared error) of out-of-sample (extrapolated) model predictions. 
When the life-stress distribution is correctly specified, we show that the optimal number of stress levels is related to the number of model parameters. In terms of test unit allocation, we show that the proportion of test units is inversely related to the stress level. Our general simulation framework provides an alternative approach to theoretical optimisation, and is particularly favourable for large/complex multipoint test plans where analytical optimisation could prove intractable.  Our procedure can be applied to a broad range of experimental scenarios, and serves as a useful tool to aid practitioners seeking to maximise component lifetime information through accelerated life testing.\\

\noindent{\bf Keywords:}  Accelerated life test, Optimised test plan, Design of experiments, Optimal design, Reliability analysis \\

\noindent{\bf MSC 2020 subject classification:}  62N05

\end{abstract}

\qquad

\newpage


\section{Introduction}
\label{sec:Intro}

The reliability of many modern components is such that it can be quite difficult to obtain failure information using standard life tests. In other words, the time-to-failure of reliable components is often so large it is not practically feasible to obtain lifetime data under normal operating conditions. Accelerated life testing (ALT) is a method of subjecting components to extreme stress to expediate, i.e. accelerate, their failure time. ALT can be used to discover failure modes that may occur in normal use, but more typically, the goal is to use the accelerated (high stress) data to predict the component lifetime under intended operating conditions (design stress) \citep{Nelson:2005}. This involves fitting a statistical model to the accelerated data from which extrapolations are made to estimate the lifetime at the design stress.

A fundamental aspect in ALT is the design of the testing experiment, the so-called accelerated life test plan. There are several parameters to consider when designing a test plan, for example, the number of stresses and their levels, the stress loading mechanism(s), the test duration, the number of items to be tested at each stress level etc. The goal is to design a testing experiment, i.e., configure the test plan parameters, in order to maximise the lifetime information gained from the testing experiment. In this work, we propose a simulation-based optimisation procedure to determine optimal accelerated life test plans for components subjected to a single (and constant) accelerating stress. 

The proposed simulation framework allows a variety of test plans to be evaluated and compared offline, thus offering a more time and cost efficient alternative to physical (destructive) component testing. By generating random lifetime data, we can compare the true (known) parameters against values estimated from statistical models for a given test plan configuration. Within this simulation framework we can modify the number of design points (stresses) and their levels, alter the test unit allocation, and explore various life-stress relationships to assess a range of test plans in terms of their design-stress lifetime prediction performance. Using the statistical computing language \texttt{R}, we optimise accelerated test plans by employing the differential evolution (DE) optimisation algorithm (and \texttt{R}'s DE package \texttt{DEoptim}), where the inputs to the objective function are the test plan parameters.  For the purpose of this paper, we use the RMSE (root mean squared error) of out-of-sample lifetime predictions as our optimisation criterion. However, the overall framework, based on data simulation and DE optimisation, is sufficiently general such that any other optimality criterion and/or experimental constraints could be included. 

The task of obtaining an optimal accelerated test plan has a wide literature, with several published works focussing on specific test scenarios, i.e., test plans for a given lifetime distribution, stress type and number of stresses. For example, a two-stress, constant-stress test plan is considered by \cite{Barton:1991}. \cite{Meeker:1975} and \cite{Nelson:1978} also consider two-stress constant-stress plans in relation to the Weibull and extreme value distributions, whilst \cite{Yang:1994} discusses a constant-stress test plan with up to four stress levels. Step-stress test plans are considered by \cite{Haghighi:2014}, and \cite{Khamis:1996} discuss a three-step step-stress test for both linear and quadratic models. More recently, \cite{Ayasse:2022} demonstrate the optimisation of an accelerated life test with two stress levels using simulated data and response surface methodology. 

In contrast to much of the existing literature, we do not focus on a specific ALT scenario,  but instead propose a general and flexible framework that is both applicable across a range of ALT optimisation problems, and compatible with various common physics-based/empirical life-stress relationships. For simpler experiments, with a limited number of design points, it is possible to mathematically optimise the test plan, and we find that our simulation results are consistent with existing analytical results in such scenarios. However, for complex multipoint designs, theoretical/analytical optimisation is more challenging, and to the best of our knowledge, such analyses are lacking in the literature. Where theoretical optimisation could prove intractable in such multipoint design scenarios, our proposed optimisation procedure can easily accommodate design schemes with an arbitrarily large number of design points. Furthermore, our flexible statistical modelling framework permits any statistical distribution (with any model functional form). Therefore, complex life-stress relationships can also be incorporated within complex design schemes. Our aim is to provide actionable insight to reliability practitioners working on real-life ALT problems. As such, the optimisation framework offers a practical alternative to theoretical analysis and serves as an effective tool for optimising complex multipoint accelerated life test plans. We also make our tool available in the form of open-source \texttt{R} code available at:  \url{https://github.com/o-mcgrath/ALT_Optimisation}

The remainder of this article is organised as follows. In Section \ref{sec:Test_Plan} we discuss test plan design considerations. Details of the modelling and simulation framework are provided in Section \ref{sec:Mod_Frame}, and our optimisation procedure is applied in some commonly-used settings (linear, quadratic and power law stress) in Section \ref{sec:Optim_Results}. In Section \ref{sec:Case_Study}, we utilise a real-world ALT case study to demonstrate the application of our optimisation procedure. Finally, the main findings and results are discussed in Section \ref{sec:Discuss}.


\section{Test Plan Considerations}
\label{sec:Test_Plan}

After conducting an accelerated life test, a statistical model is fitted to the (high stress) data from which extrapolations are made to predict the lifetime at the design stress (intended operating stress). The purpose of the test plan is to facilitate the model fitting process by extracting as much information as possible from the experiment. In designing the test plan, the main considerations include: (i) accelerating variable(s), (ii) stress loading, (iii) test duration, (iv) lifetime distribution, (v) life-stress relationship and (vi) optimality criterion. The extensive reviews from \cite{Chen:2018}, \cite{Limon:2017} and \cite{Nelson:2005} draw the conclusion that the majority of test plans utilise a single (constant stress) accelerating variable and Type I data where components are tested until the earlier of failure or a specified testing duration, and an assumed Weibull lifetime distribution.

We aim to provide a pragmatic optimisation framework that supplements physical ALT experiments. Therefore, in this work, we focus on the experimental framework most likely to be adopted in practice: Type I data, single-variable, constant-stress test plans. In relation to the optimality criterion, most works aim to minimise the asymptotic variance of the maximum likelihood estimate (MLE) of an estimated quantile for a specified low stress \citep{Nelson:2005, Chen:2018}. In this work, we consider both estimation variability and bias using the RMSE of the estimated quantile, where we numerically approximate the RMSE through simulation. In terms of specifying a lifetime distribution, we align with the ALT literature, and focus on the Weibull distribution. However, as outlined in the following section, our modelling framework is sufficiently general to accommodate models with any underlying lifetime distribution, and with any lifetime-stress relationship.


\section{Modelling Framework}
\label{sec:Mod_Frame}

Estimating the design-stress lifetime requires a statistical model that faithfully describes the underlying lifetime-stress relationship. Once such a model is obtained, one can extrapolate the model to predict the lifetime at the design stress. The optimal test plan is that which maximises the lifetime information, ultimately improving the prediction performance of the fitted model. In this section we detail the statistical modelling framework and outline our simulation-based approach to obtaining optimal accelerated test plans.


\subsection{Statistical Model}
\label{sec:Stat_Model}

Denoting $T\in [0, \infty)$ as the time-to-failure, the probability that a component survives (does not fail) beyond time $t$ is given by the reliability function
\be R(t) = \Pr(T > t) = \displaystyle \int _t^\infty \hspace{-0.15cm} f(u) \hspace{0.05cm} du,\ee where $f(u)$ is the probability density function.

In this work, we consider Type I censored lifetime data, i.e., components tested until failure, or for a specified duration $t_e$, whichever comes first. If components do not fail during the experiment, they are said to be right censored, and provide only partial information, i.e., $T \ge t_e$.  Letting $\delta_i \in \{0,1\}$ be an indicator function, where $\delta_i = 1$ if lifetime $t_i$ is uncensored, and $\delta_i = 0$ if lifetime $t_i$ is censored, the right-censored log-likelihood function is given as
\be l(\theta) = \displaystyle\sum_{i=1}^{n} \delta_i\log f(t_i) + (1-\delta_i)\log R(t_i),\ee where $\theta$ is a vector of parameters describing the probability distribution, and $n$ is the total number of components tested. Denoting the accelerating stress variable as $S$, we can obtain the stress-dependent reliability function $R(t|S=s)$, and calculate quantiles of interest from
\be q_\tau(s) = R^{-1}(1-\tau|S=s).\ee Letting $s_d$ be the design stress, in this paper, we focus on obtaining the median design-stress lifetime, $q_{0.5}(s_d)$. However, the proposed framework permits any quantile (or indeed any quantity) of interest to be considered.

In ALT, a life-stress relationship is often specified from empirical or physics-based models. For example, when temperature is used as the stress condition the Arrhenius relationship can be used \citep{Ayasse:2022, Limon:2017}. When the accelerating variable is voltage, the inverse power law model is often used \citep{Barton:1991, Meeker:1975}, with the generalised Eyring model being a popular choice to capture the combined effects of thermal and non-thermal stresses \citep{Limon:2017}. However, in this work, we adopt a very general modelling approach, utilising an \emph{accelerated failure time} (AFT) model of the form \begin{equation} \label{eq:loglin} \log T(S) = \mu(S) + \sigma \epsilon,\end{equation}
where the log-link is used to ensure positivity of lifetime predictions, $\mu$ and $\sigma$ are the location and scale parameters (respectively), and $\epsilon$ is an error distribution. This general framework captures many popular physics-based models such as the thermochemical model (via $\mu(S) = \beta_0 + \beta_1 S$), Arrhenius model (via $\mu(S) = \beta_0 + \beta_1 / S$), power law model (via $\mu(S) = \beta_0 + \beta_1 \log (S)$) and the exponential $\sqrt{S}$ model (via $\mu(S) = \beta_0 + \beta_1 \sqrt{S}$) as shown in Appendix \ref{App:A}. The framework could also be used for more purely ``empirical'' modelling such as polynomial life-stress relationships, e.g., via $\mu(S) = \beta_0 + \beta_1 S + \beta_2 S^2$.

The comprehensive bibliographies of \cite{Chen:2018} and \cite{Limon:2017} show that in terms of specifying a life-stress relationship, the majority of references use the exponential, log-normal and Weibull distributions. Although the modelling framework outlined above can easily accommodate any probability distribution (through choice of the distribution of $\epsilon$), in this paper, we focus on the most widely used distribution, the Weibull. Given $T\sim \text{Weibull}$, the median design-stress lifetime is given as \be \label{eq:medpred}q_{0.5}(s_d) = R^{-1}(0.5|S=s_d)= \exp(\mu(s_d))\log(2)^{\sigma(s_d)}.\ee 


\subsection{\textit{Optimisation Framework}}
\label{sec:Optim_Frame}

Denoting $N$ as the number of stresses (design points), $\bm{S}= \{S_1, S_2, \cdots, S_N\}$ as the stress levels, $D$ as the minimum allowable distance between consecutive stresses, $\bm{P} = \{p_1, p_2, \cdots, p_N\}$ as the proportion of test units allocated to a given stress level ($p_1 + p_2 + \cdots + p_N = 1$), $t_e\in(0,\infty)$ as the experiment duration and $Q= q_{0.5}(s_d)$ as the median lifetime at the design stress, the objective function of interest (which we aim to minimise) is given by
\be \text{RMSE}(N, \bm{S}, D, \bm{P}, t_e) = \left(E[\hat{Q} - Q]^2\right)^{1/2}, \ee 

where $\hat{Q}$ is an estimate of $Q$.  This RMSE objective function depends on the configuration of the testing experiment, i.e., the test plan parameters, $N$, $D$, $\bm{S}$, $\bm{P}$ and  $t_e$, as a direct consequence of $\hat Q$ depending on the test plan, whereas $Q$ is a fixed constant. In general, particularly for complex design scenarios, it may prove infeasible to compute the RMSE analytically. Therefore, in order to provide a sufficiently general optimisation framework that accommodates complex designs, we propose approximating the RMSE by way of Monte Carlo simulation, i.e., \bne \label{Eq:RMSE_sim} \text{RMSE}(N, \bm{S}, D, \bm{P}, t_e) \approx \left(\dfrac{1}{n_\text{sim}}\displaystyle \sum _{j=1} ^ {n_\text{sim}}[\hat{Q_j} - Q]^2\right)^{1/2}, \ene where $n_\text{sim}$ is the number of simulation replicates. Of course, the larger the value of $n_\text{sim}$, the lower the Monte Carlo error, i.e., the closer (\ref{Eq:RMSE_sim}) will be to the analytic (but intractable) RMSE. In this paper, we use $n_\text{sim} = 1000$, but the results are very similar for $n_\text{sim} = 500$ (see Appendix \ref{App:B_nsim}).

Given that the RMSE objective depends on the inputs ($N, \bm{S}, D, \bm{P}, t_e$) in a complex non-linear (and possibly non-differentiable) manner, we use the differential evolution (DE) optimisation algorithm. The DE algorithm is a type of genetic algorithm that adopts a biological, mutation-based approach to evolve the objective function over successive iterations (generations). (See \cite{Storn:1997} for further details.) Within our optimisation framework, we specifically utilise the DE algorithm from the \texttt{R} package \texttt{DEoptim}, developed by \cite{Mullen:2011}. The DE algorithm has been shown to perform well across a wide range of situations where the objective function is noisy/discontinuous and/or difficult to differentiate \citep{Mullen:2011}. Additionally, \cite{Price:2005} demonstrate the versatility (and robustness) of the algorithm in the context of various practical problems. Thus, the DE algorithm, and its \texttt{R} implementation, \texttt{DEoptim}, is an appropriate choice for our general modelling and test-plan setup with stochastically-approximated RMSE objective function. It is clear that our proposal is quite flexible: (i) an objective function of interest is computed, which could be RMSE (per this paper) or some other function; (ii) the computation is based on simulated data, wherein any realistic experimental constraints can be included, e.g., the experimental duration, the minimum distance between stresses, the stresses having to fall in a discrete set, or the number of test units having to be multiples of some number; and (iii) given the objective function and experimental constraints, the design is optimised using the general purpose DE algorithm.

It is important to note that our proposed model fitting/optimisation framework relies on generating random accelerated data. In practice, when generating lifetime data, an analyst must choose simulation parameters that are compatible with the parameters of the true (perhaps unknown) life-stress distribution. In the following section, we provide a brief discussion of this challenge, and consider possible approaches to overcome it. However, note that our article is primarily focused on a general framework for producing an optimal test plan \emph{given} model parameters.


\subsection{\textit{Parameter Pre-Estimation}}
\label{sec:Param_Pre}

As mentioned in the previous section, to produce an optimal test plan in practice, an initial estimate of model parameters is needed in order to generate accelerated data that is consistent with the underlying component lifetime distribution. This problem of requiring knowledge of the true model parameters is inherent in \emph{all} accelerated life test plan optimisation \citep{Nelson:1978, Nelson:2005}. \cite{Chen:2018} also acknowledge that initial model parameter estimates are required to design an optimal ALT test plan. One solution is to use a Bayesian analysis approach where unknown values are replaced with prior distributions. Examples of Bayesian approaches in ALT planning include \cite{Chaloner:1992}, \cite {Naijun:2017} and \cite{Zhang:2006}. 

Another common approach is to conduct a pilot study/preliminary test and/or utilise lifetime information from a similar component. For example, in their discussion of optimising a two-stress temperature-stress test plan, \cite{Ayasse:2022} use historical data from a similar product to obtain log-normal model parameter estimates. \cite{Barton:1991} similarly provides a temperature-based example (using the Arrhenius model) that utilises preliminary test data. In a discussion on assumed parameter values, \cite{Nelson:1978} state that, in practice, parameters must be estimated using experience, similar data, or a preliminary test. \cite{Yang:1994} too discusses the issue of pre-estimating unknown parameters and suggests the same solutions as \cite{Nelson:1978}, but also mentions the Bayes approach. In this work, we adopt the practical and commonly-used approach of obtaining initial parameter estimates from a preliminary testing experiment. 


\subsection{\textit{Preliminary Test Data}}
\label{sec:Pre_Data}

Our approach is to fit a suitable reliability model to preliminary test data, and use the fitted model coefficients as the parameters in the optimisation. We consider a preliminary test where $n$ components, each with time-to-failure, $T$, are subjected to a single (constant) accelerating stress, $S \in (0,1)$, such that $T|(S=0) = \infty$, and $T|(S=1) = 0$. In practice, a preliminary experiment may have physical constraints such as testing equipment capacity, component availability etc., and as such, will be application specific.  For demonstration purposes, in the next section, we consider $n=180$ components tested at the following stresses: $\bm{S} = \{0.1, 0.2, \cdots, 0.8, 0.9\}$, and specify a design stress of $s_d=0.05$. In standard constant-stress tests, it is common to use equally spaced stresses, with equal test unit allocation \citep{Tang:2002}, so for this initial test, we consider a balanced design, i.e., $n_1 =n_2\cdots = n_9 = 20$. 
In this paper, we focus on preliminary test data which follows a Weibull distribution, and consider some common statistical functional forms, specifically: linear, quadratic and power law relationships.  In the following section, we give a brief overview of the optimisation schemes considered, and discuss the results of each optimisation in detail.


\section{Optimisation Results}
\label{sec:Optim_Results}
For each model functional form, we consider the following optimisation strategies:\\(i) For each fixed $N \in \{2,3,4,5,6\}$, determine the optimal $N$-point design. $N$ is specified and the objective function inputs are the stress levels, $\bm{S}$, the minimum distance between consecutive stresses, $D$, the test units at each stress, given by $n \times \bm{P}$ and the experiment duration, $t_e$. (Of course, larger values of $N$ can be used within our framework.)\\
(ii) Consider the more general setting where $N$ is an objective function input as per (\ref{Eq:RMSE_sim}).\\
(iii) Compare the optimal test plan with a number of similar plans. This provides some comparable test plans which could be used in scenarios where the optimal test plan cannot be implemented (due to testing/equipment constraints, for example).

We apply each of the above optimisation/analysis strategies to data assuming different life-stress shapes, namely, linear, quadratic and power law. For each strategy and life-stress functional form, we apply our \texttt{DEoptim} optimisation framework and discuss the resulting optimal test plan configurations.

Note: In our analysis, the number of optimisation iterations (\texttt{DEoptim} generations) is fixed at 200 (\texttt{DEoptim} default), however, this parameter is easily adjusted within the optimisation framework. It is largely a matter of computational resources and time-sensitivity of results: it could be set smaller if fast initial plans are needed (e.g., in some initial exploratory work) or larger for a final optimisation.


\subsection{\textit{Linear Life-Stress Relationship}}
\label{sec:Linear}

\begin{figure}[!h]
\centering
\includegraphics[width=11cm, height=8cm]{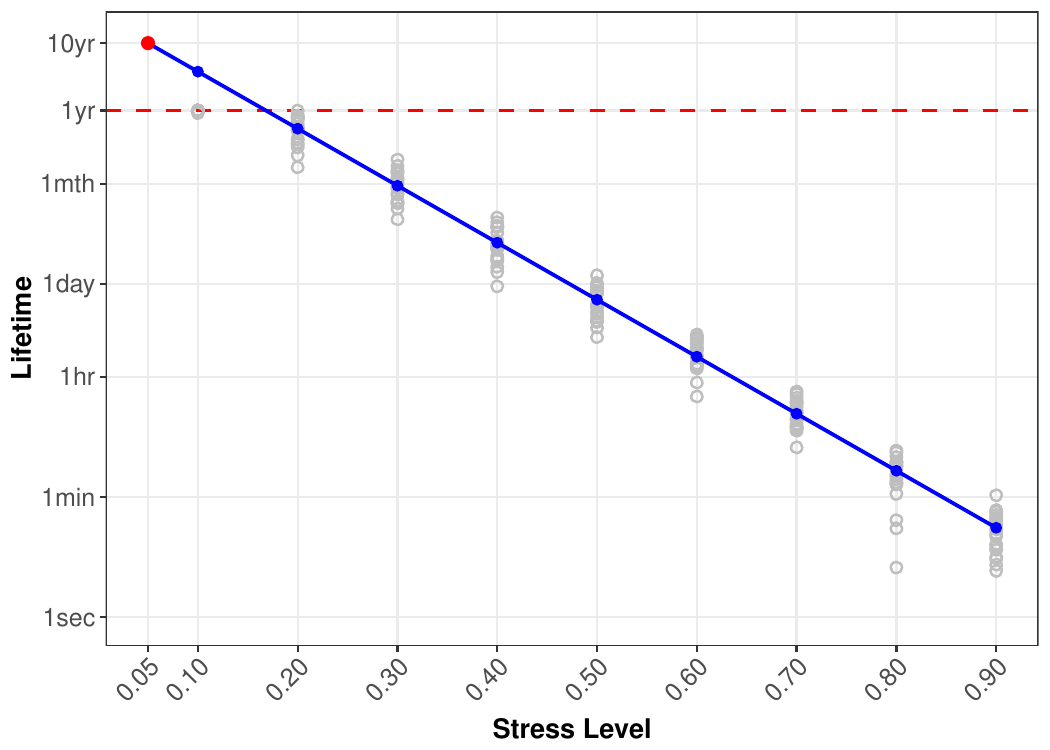}
\caption{Sample preliminary ALT data depicting a linear trend. Horizontal (red) dashed line shows the experiment duration, $t_e$. Component lifetimes are shown as (grey) open circles. Censoring occurs at $S=0.1$ and $S=0.2$ ($95\%$ and $5\%$ right-censoring respectively). Estimated lifetime at the design stress ($S=0.05$) is shown as solid (red) point. }
\label{Fig:data_linear}
\end{figure}

Consider conducting a pilot study/preliminary ALT experiment, and obtaining a linear life-stress relationship as shown in Figure \ref{Fig:data_linear}. 

We fit a linear AFT model, i.e., $\mu(S)= \beta_0 + \beta_1S$, and use the fitted model parameters ($\hat{\beta_0} = 12.5, \hat{\beta_1} = -19.5, \hat{\sigma} = 0.5$) as the basis of the optimisation.  Of course, the validity of the subsequent optimisation depends on sensible initial parameter estimates, i.e., fitting a reasonable model to the preliminary data. Therefore, one could consider utilising statistical model selection procedures in this initial phase, but here, the focus is on optimising the test plan (given an initial model has been obtained). 

We consider obtaining individual optimal test plans for $N$-point designs where $N \in\{2,3,4,5,6\}$, and also consider a separate  analysis where the optimiser additionally optimises $N$. For each optimisation, the stress level range is $\bm{S}=[0.10, 0.90]$ with a minimum stress distance (granularity) $D$, such that $S_{j+1} \ge S_{j} + D$. In this work, we specify a small value of $D = 10^{-5}$, which, from a practical perspective, essentially allows a given $N$-point design the ability to reduce to a lower-than-$N$-point design. For example, for the $N=6$ design shown in Table \ref{Tab:results_linear}, stresses $S_3, S_4, S_5$ and $S_6$ are each separated by a distance of $D=10^{-5}$ which, from a practical perspective (rounding to two decimal places), can be considered as a single stress $S_3=0.90$. In some situations this property may not be desirable, for example, a practitioner might need to more strictly enforce an $N$-point design, in which case $D$ can set to be a larger number, e.g., to align with the stress resolution of their physical experimental setup. In Appendix \ref{App:B_stress_distance}, we provide optimisation results with the constraint that $D=10^{-1}$, which then creates a bigger separation between neighbouring stress values.

In terms of experimental test units, we consider a total of $n=100$ in the optimisation, where the minimum number of test units at a given stress level is $n_j=1$. Note that $n$ is a relatively arbitrary number here as it can always be scaled up to reduce the RMSE. Indeed, for the linear AFT scenario, the RMSE is reduced by approximately $30\%$ when the number of test units is increased from $n=100$ to $n=200$; this is shown in Appendix \ref{App:B_test_units}.

We remind the reader of the stochastic nature of the overall optimisation framework: \texttt{DEoptim} is a stochastic optimiser, which is minimising a stochastic objective function (i.e., Monte-Carlo-calculated RMSE). Therefore, the test plan results will naturally vary (slightly) from one run of the algorithm to another. To account for this variability, in addition to reporting the minimum RMSE obtained by \texttt{DEoptim}, we also report the mean, standard error, and the 5\textsuperscript{th} and 95\textsuperscript{th} percentiles of the RMSE following 1000 simulation replicates at the given \texttt{DEoptim} test plan. These additional metrics are important for understanding whether or not the given plans differ statistically from each other. (Of course, one can also run \texttt{DEoptim} itself multiple times, and we have run some variations of the linear-AFT setup provided in Appendix \ref{App:B}, all of which display similar results to Table \ref{Tab:results_linear} in terms of best test plan.)

\begin{table}[!h]
\small
\centering
\caption{Optimal test plan configurations for linear AFT model. For this and subsequent tables, stress levels are printed correct to 2 decimal places. Values in parentheses show the associated number of test units. $N=6^*$ refers to the optimisation where $N$ is an objective function input. The lowest RMSE values are shown in bold.}
\label{Tab:results_linear}
\begin{tabular}{c|cccccc|c|c|c|c|c} 

\multirow{2}{*}{$N$} & \multicolumn{6}{c|}{Stress Levels}   &\multicolumn{1}{c|}{Min.} &\multicolumn{1}{c|}{Mean} &  \multicolumn{1}{c|}{Std.} & \multirow{2}{*}{$p_{0.05}$} & \multirow{2}{*}{$p_{0.95}$} \\
 & $S_1$ & $S_2$ & $S_3$ & $S_4$ & $S_5$ & $S_6$ & \multicolumn{1}{c|}{RMSE} & \multicolumn{1}{c|}{RMSE} & \multicolumn{1}{c|}{Error} &&  \\
\hline
&&&&&&&&&&\\[-0.45cm]
\multirow{2}{*}{2} & 0.19 & 0.89 & \multirow{2}{*}{---} & \multirow{2}{*}{---} & \multirow{2}{*}{---} & \multirow{2}{*}{---} & \multirow{2}{*}{6,446}  & \multirow{2}{*}{\textbf{7,038}}& \multirow{2}{*}{169}& \multirow{2}{*}{6,767} & \multirow{2}{*}{7,317}\\
&(85)&(15)&&&&&&&&&\\
&&&&&&&&&&&\\[-0.45cm]
\hline
&&&&&&&&&&&\\[-0.45cm]
\multirow{2}{*}{3} & 0.20& 0.90 & 0.90 & \multirow{2}{*}{---} & \multirow{2}{*}{---}& \multirow{2}{*}{---}  & \multirow{2}{*}{6,532} & \multirow{2}{*}{7,082}& \multirow{2}{*}{165} & \multirow{2}{*}{6,812} & \multirow{2}{*}{7,358}\\
&(85) & (01) & (14) &&&&&&&&\\
&&&&&&&&&&&\\[-0.45cm]
\hline
&&&&&&&&&&&\\[-0.45cm]
\multirow{2}{*}{4} & 0.19 & 0.90 & 0.90 & 0.90 & \multirow{2}{*}{---} & \multirow{2}{*}{---}  & \multirow{2}{*} {\textbf{6,404}} & \multirow{2}{*}{7,080} & \multirow{2}{*}{165} & \multirow{2}{*}{6,802} & \multirow{2}{*}{7,350} \\
&(86) & (02) & (09) & (03) &&&&&&&\\
&&&&&&&&&&&\\[-0.45cm]
\hline
&&&&&&&&&&&\\[-0.45cm]
\multirow{2}{*}{5} & 0.19 & 0.21 & 0.88 & 0.90 & 0.90 & \multirow{2}{*}{---}  & \multirow{2}{*}{6,547} &\multirow{2}{*}{7,075}& \multirow{2}{*}{164} & \multirow{2}{*}{6,798} & \multirow{2}{*}{7,335}\\
&(76) & (10) & (03) & (05) & (06) &&&&&&\\
&&&&&&&&&&&\\[-0.45cm]
\hline
&&&&&&&&&&&\\[-0.45cm]
\multirow{2}{*}{6} & 0.19 & 0.43& 0.90 & 0.90 & 0.90 & 0.90  & \multirow{2}{*}{6,543}  & \multirow{2}{*}{7,067}&\multirow{2}{*}{163} & \multirow{2}{*}{6,806} & \multirow{2}{*}{7,343}\\
&(84) & (01) & (01) & (07) & (02) & (05) &&&&& \\
&&&&&&&&&&&\\[-0.45cm]
\hline
&&&&&&&&&&&\\[-0.45cm]
\multirow{2}{*}{6\textsuperscript{*}} & 0.19 & 0.27 & 0.90 & 0.90  & 0.90 & 0.90   & \multirow{2}{*}{6,455} & \multirow{2}{*}{7,070}& \multirow{2}{*}{166} & \multirow{2}{*}{6,788} & \multirow{2}{*}{7,353}\\
&(82) & (01) & (01) & (01) & (01) & (14)  & & &&&\\
\end{tabular}
\end{table}

The optimal test plans pertaining to the linear test data (Figure \ref{Fig:data_linear}) are listed in Table \ref{Tab:results_linear}, indicating that a 2-point design is best. In particular, the 2-point design has the lowest mean RMSE (7,038), and suggests $n_1 = 85$ units at stress $S_1 = 0.19$ and $n_2 = 15$ units at stress $S_2 = 0.89$. Given the values of standard errors and percentiles, it is clear that this design does not differ statistically from the other $N$-point designs. Moreover, given the small $D = 10^{-5}$ value we have used, from a practical perspective, all designs essentially collapse to a 2-point design where most test units are assigned to the low stress (of around 0.2) with the remaining units assigned to the high stress (of around 0.9). 

As per Figure \ref{Fig:data_linear}, $95\%$ of the lifetimes at the minimum stress ($S=0.1$) are right-censored, with much lower censoring ($5\%$) occurring at $S=0.2$. Thus, for this linear setting, it is best to test more units at a lower stress where there is some (but not much) censoring, and fewer units at a much higher stress where there is more information per unit tested (i.e., at the higher stress, units fail much quicker and are more concentrated around the mean lifetime). 

\begin{figure}[!h]
\centering
\includegraphics[width=11cm, height=8cm]{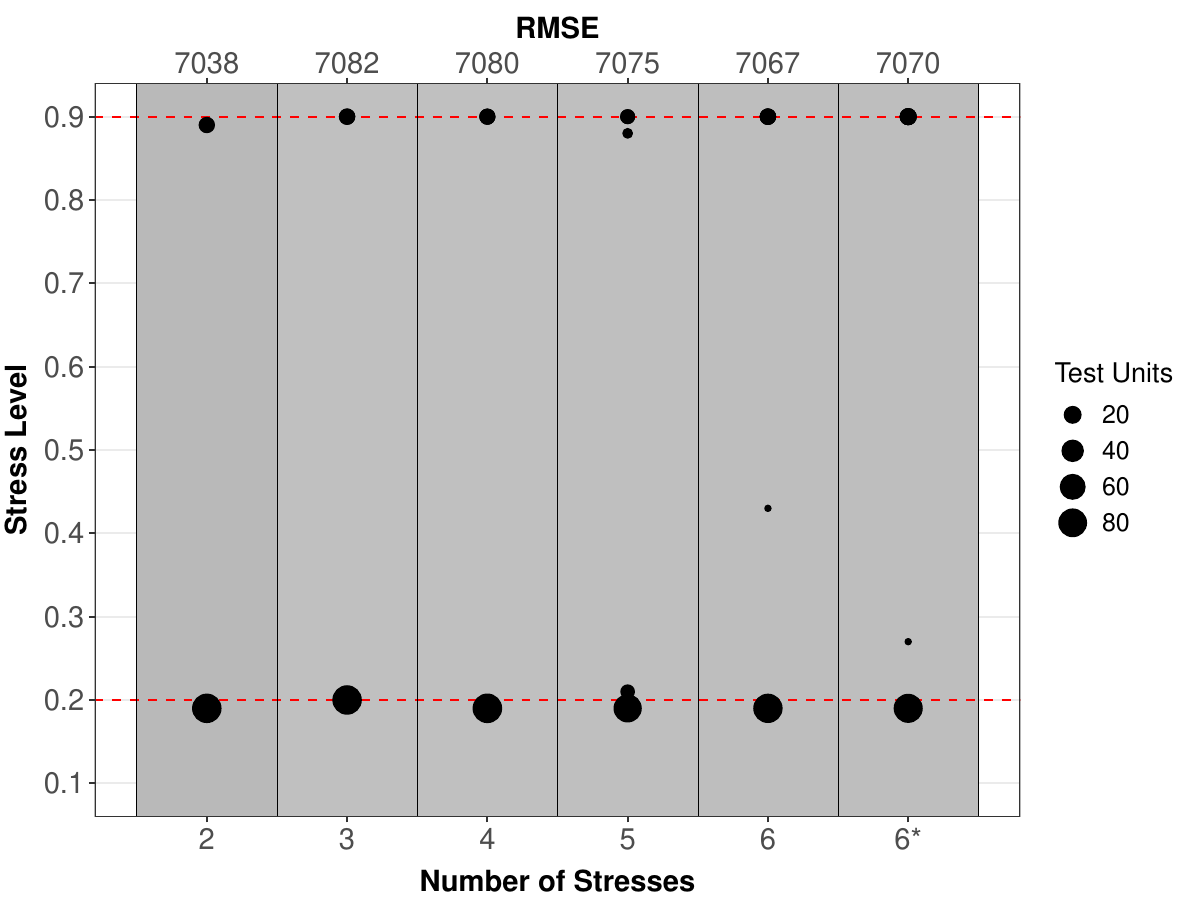}
\caption{Comparison of optimal test plans and RMSE for linear AFT model.  Solid (black) data points show the stress levels. The size of the points are proportional to the test unit allocation. RMSE values for each test plan are compared using greyscale shading, where light-to-dark shading (white-to-black) corresponds to the smallest mean RMSE plus/minus three times the associated standard error. (But all designs end up with the same shade of grey as they differ so little statistically.)  }
\label{Fig:Three_in_one_linear}
\end{figure}

Figure \ref{Fig:Three_in_one_linear} provides a visualisation of the results of Table \ref{Tab:results_linear}, where for each $N$, the optimiser converges to $S_1\approx0.2$, $S_N\approx0.9$, with a higher proportion of test units allocated to the lower stress. For $N \in \{5,6,6^*\}$, we note that the optimiser has allocated a very small number of units to other stresses. Given the complex and stochastic nature of the optimisation problem, it is likely that the optimiser has not yet fully converged to a 2-point design within the given number of iterations. From a practical perspective of physical testing, it is unlikely that one would run a given design point with such a small number of units, when not supported by any meaningful improvement in RMSE. More precisely, reallocating these ``spurious'' few units to the smaller and larger stresses collapses to the 2-point design, which yields the same RMSE from a statistical perspective (as indicated by the grey shading in the plot).\\

Next, we assess the performance of some practical variations of the optimal test plan. This provides practitioners with insights into some alternative test plans which, perhaps due to real experimental constraints, might be more convenient to implement. (It is worth noting that such real-world physical constraints can also be embedded directly into our optimiser as the data can be simulated subject to any constraints.)

\begin{figure}[!h]
\centering
\includegraphics[width=11cm, height=8cm]{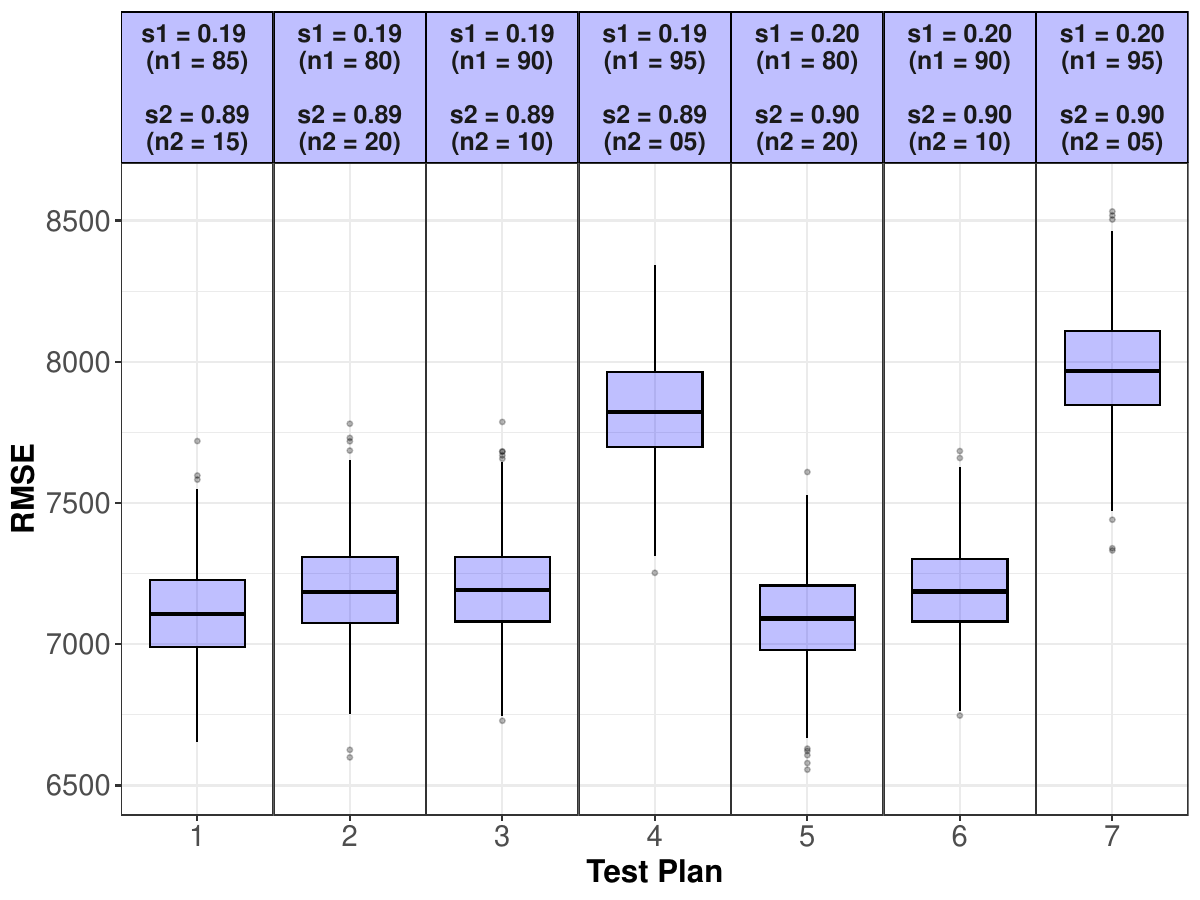}
\caption{Comparison of 2-point linear test plans with various test unit allocations and/or rounding of stresses. Test plan \# 1 is the optimal test plan.}
\label{Fig:compare_linear}
\end{figure}

A comparison of variations of the optimal test plan is given in Figure \ref{Fig:compare_linear}. Here we compare the optimal test plan ($S_1=0.19$, $n_1=85$; $S_2=0.89$, $n_2=15$) against several test plans where the test units are rounded (plans \#2, \#3 and \#4), and stresses are also rounded (plans \#5, \#6 and \#7). Using the optimal stresses, when the test units are within $\pm5$ of the optimal allocation (test plans \#2 and \#3), the performance does not differ much from the optimal test plan (plan \#1). However, when the test units deviate by $\pm10$ from the optimal, the RMSE does increase significantly. Referring to the test plans where both test units and stresses are rounded, the average RMSE for test plans \#5 and \#6 are very similar to the optimal plan. However, as with test plan \#4, the RMSE of test plan \#7 is larger. This analysis shows that we can perturb the test plan slightly (to accommodate practical constraints) without any significant loss of performance.

Overall, the optimisation results indicate that for Weibull lifetime data with a linear life-stress relationship, the optimal test plan configuration is to use two stress levels: the highest stress, and a low stress at which almost all failures are observed (i.e., there is very little censoring). In terms of test unit allocation, the optimal strategy is to allocate most units (approximately $85\%$) to the lower stress.  These results are consistent with the literature. (See \cite{Meeker:1975}, \cite{Barton:1991} and \cite{Nelson:2005}, for example.)\\
 
In the following sections, we conduct a similar analysis using preliminary test data that assumes quadratic and power law trends. Once the estimated coefficients are obtained, we apply the optimisation procedure and assess the results as outlined in this section.


\subsection{\textit{Quadratic Life-Stress Relationship}}
\label{sec:Quad}

\begin{figure}[!h]
\centering
\includegraphics[width=11cm, height=8cm]{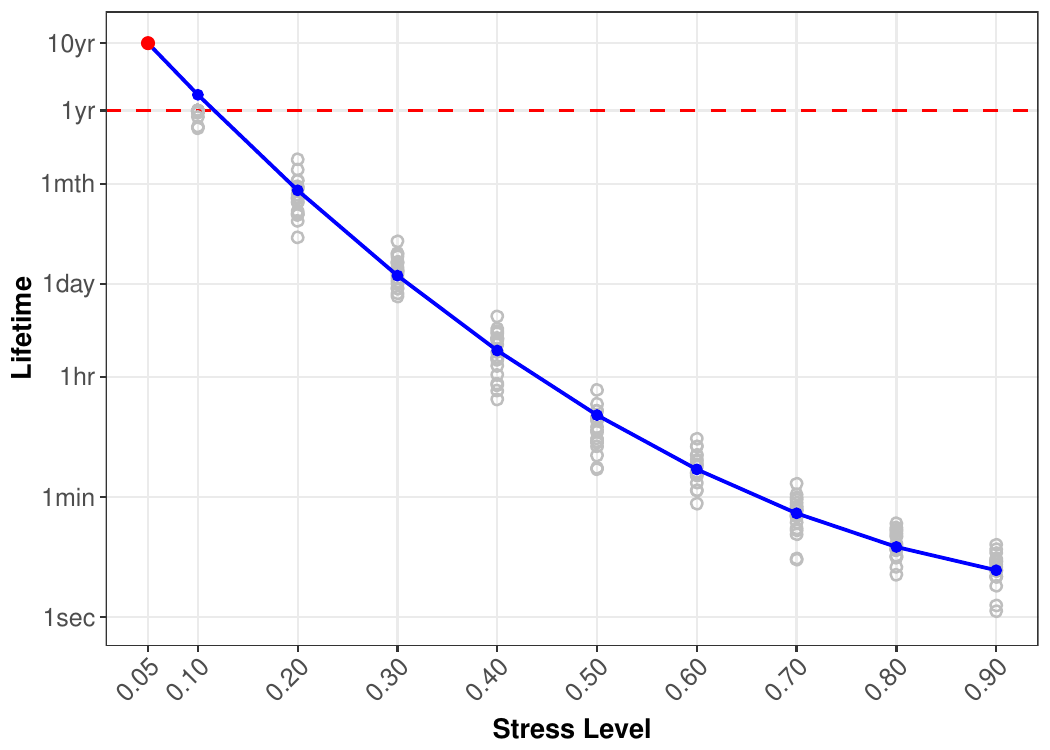}
\caption{Sample preliminary accelerated test data depicting a quadratic trend. Censoring occurs at $S=0.1$ ($80\%$ of lifetimes are right-censored). }

\label{Fig:data_quadratic}
\end{figure}

The preliminary test data given in Figure \ref{Fig:data_quadratic} shows a quadratic life-stress relationship, hence we fit a model of the form: $\mu(S) = \beta_0 + \beta_1S + \beta_2S^2$. (The estimated model coefficients are: $\hat{\beta_0} = 13.4, \hat{\beta_1} = -37.9, \hat{\beta_2} = 17.7, \hat{\sigma} = 0.5$.)\\

\vspace{-0.5cm}
\begin{table}[!h]
\small
\centering
\caption{Optimal test plan configurations for quadratic AFT model.}
\label{Tab:results_quadratic}
\begin{tabular}{c|cccccc|c|c|c|c|c} 

\multirow{2}{*}{$N$} & \multicolumn{6}{c|}{Stress Levels}   &\multicolumn{1}{c|}{Min.} &\multicolumn{1}{c|}{Mean} &  \multicolumn{1}{c|}{Std.} & \multirow{2}{*}{$p_{0.05}$} & \multirow{2}{*}{$p_{0.95}$} \\
 & $S_1$ & $S_2$ & $S_3$ & $S_4$ & $S_5$ & $S_6$ & \multicolumn{1}{c|}{RMSE} & \multicolumn{1}{c|}{RMSE} & \multicolumn{1}{c|}{Error} &&  \\
\hline
&&&&&&&&&&\\[-0.45cm]
\multirow{2}{*}{3} & 0.13 & 0.49 & 0.90 & \multirow{2}{*}{---} & \multirow{2}{*}{---}& \multirow{2}{*}{---}  & \multirow{2}{*}{8,275} &  \multirow{2}{*}{9,004}& \multirow{2}{*}{211} & \multirow{2}{*}{8,670} & \multirow{2}{*}{9,347}\\
&(66) & (27) & (07) &&&&&&&&\\
&&&&&&&&&&&\\[-0.45cm]
\hline
&&&&&&&&&&&\\[-0.45cm]
\multirow{2}{*}{4} & 0.13 & 0.52& 0.90 & 0.90 & \multirow{2}{*}{---} & \multirow{2}{*}{---}  & \multirow{2}{*} {8,411}  & \multirow{2}{*}{\textbf{8,910}} & \multirow{2}{*}{210} & \multirow{2}{*}{8,568} & \multirow{2}{*}{9,268} \\
&(71) & (24) & (04) & (01) &&&&&&&\\
&&&&&&&&&&&\\[-0.45cm]
\hline
&&&&&&&&&&&\\[-0.45cm]
\multirow{2}{*}{5} & 0.13 & 0.51 & 0.89 & 0.90 & 0.90 & \multirow{2}{*}{---}  & \multirow{2}{*}{8,421} &\multirow{2}{*}{8,979}& \multirow{2}{*}{212} & \multirow{2}{*}{8,645} & \multirow{2}{*}{9,336}\\
&(69) & (25) & (02) & (03) & (01) &&&&&&\\
&&&&&&&&&&&\\[-0.45cm]
\hline
&&&&&&&&&&&\\[-0.45cm]
\multirow{2}{*}{6} & 0.13 & 0.53 & 0.62 & 0.90 & 0.90 & 0.90  & \multirow{2}{*}{8,288}  & \multirow{2}{*}{8,918}&\multirow{2}{*}{213} & \multirow{2}{*}{8,589} & \multirow{2}{*}{9,275}\\
&(71) & (21) & (01) & (03) & (01) & (03) &&&&& \\
&&&&&&&&&&&\\[-0.45cm]
\hline
&&&&&&&&&&&\\[-0.45cm]
\multirow{2}{*}{6\textsuperscript{*}} & 0.13 & 0.53 & 0.58  & 0.90   & 0.90  & \multirow{2}{*}{---}    & \multirow{2}{*}{\textbf{8,239}} &  \multirow{2}{*}{9,062}& \multirow{2}{*}{219} & \multirow{2}{*}{8,717} & \multirow{2}{*}{9,429}\\
&(75) & (16) & (01)   & (01)  & (07)   &    & & &&&\\
\end{tabular}
\end{table}

\begin{figure}[!h]
\centering
\includegraphics[width=11cm, height=8cm]{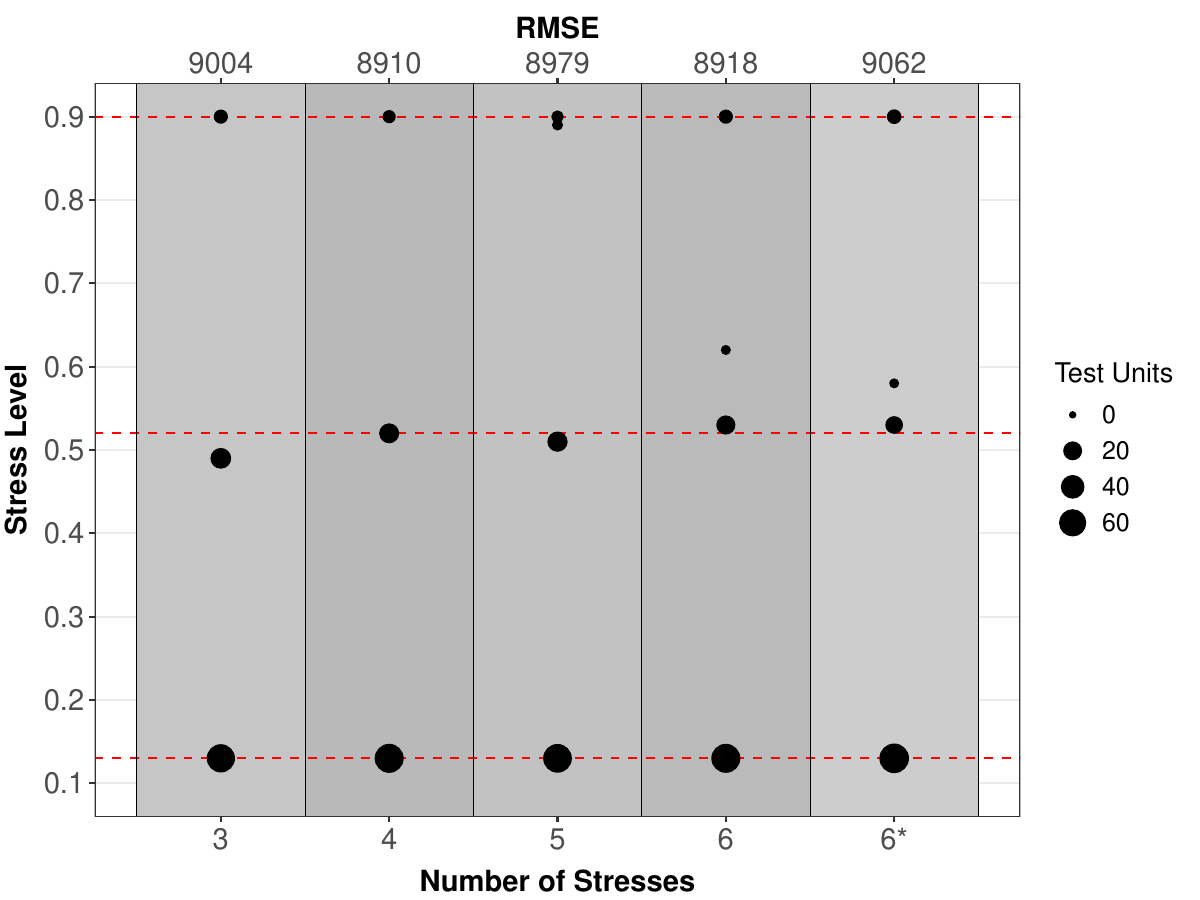}
\caption{Comparison of optimal test plans and RMSE for quadratic AFT model; greyscale colouring as per Figure \ref{Fig:Three_in_one_linear}.}
\label{Fig:Three_in_one_quadratic}
\end{figure}

Table \ref{Tab:results_quadratic} shows that the optimiser converges to a 3-point design: $S_1=0.13$, $S_2\approx0.5$, $S_3=0.9$. Akin to the linear test plans, the test unit allocation is inversely related to stress level, i.e., the larger the stress, the fewer units allocated. We can see that the test plans are very similar: for each $N$, the stress levels are reasonably consistent with relatively small variation in test unit allocation. As shown in Figure \ref{Fig:Three_in_one_quadratic}, each test plan exhibits a similar grey shade indicating that there is little statistical difference between test plans. (Recall that the greyscale shading is based on the standard error of the RMSE.) For $N=\{5,6,6\textsuperscript{*}\}$, the optimiser identifies a fourth stress with a small number of test units ($n \le 2$). Given the small number of test units, it seems reasonable (from the point of view of physical testing), to consider these as 3-point plans. Overall, given that the test plans are not statistically different, the optimisation results indicate that a test plan with three stresses is optimal.

\begin{figure}[!h]
\centering
\includegraphics[width=11cm, height=8cm]{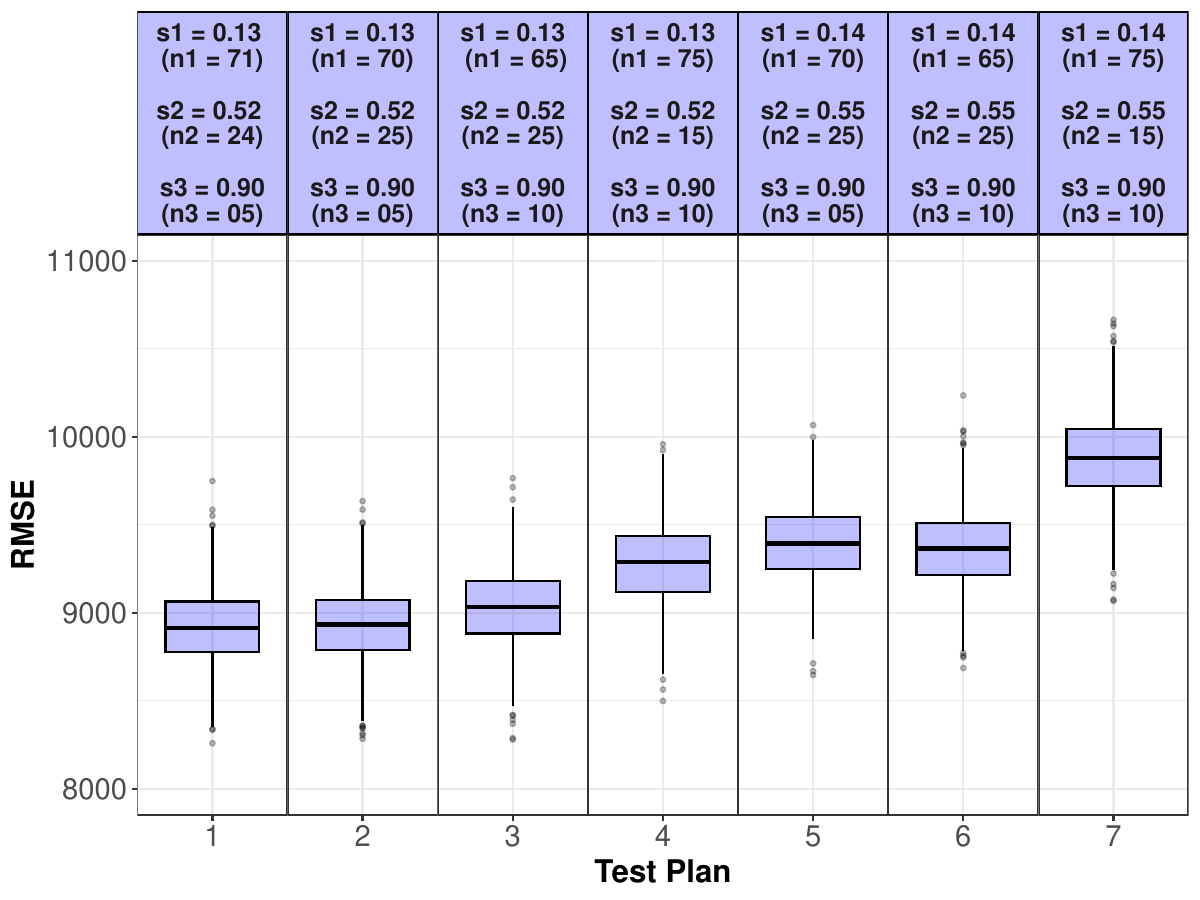}
\caption{Comparison of 3-point quadratic test plans. (Test plan \# 1 is the optimal plan.)}
\label{Fig:compare_quadratic}
\end{figure}

A comparison of 3-point quadratic test plans is given in Figure \ref{Fig:compare_quadratic}. Test plans \#2, \#3 and \#4 have rounded test units, and test plans \#5, \#6 and \#7 have both test units and stresses rounded. We can see that slight rounding of test units (plans \#2 and \#3) provides comparable performance to the optimal plan (plan \#1), whereas the other deviations presented here (in plans \#4, \#5, \#6, \#7) lead to larger increases in RMSE.

Overall, in relation to a quadratic Weibull AFT model, the optimal test plan configuration is to use three stress levels: a low stress, an intermediate stress, and a high stress. This coincides with previous works in the literature, for example: \cite{Khamis:1996} and \cite{Yang:1994}. In relation to test units, the optimal strategy is to allocate more test units to lower stresses; again aligning with the literature (see \cite{Nelson:2005}.) The test unit allocation shown in Table \ref{Tab:results_quadratic} is in line with the common rule of thumb to allocate units to the low, medium and high stresses in the ratio 4:2:1 \citep{Zhu:2013}. More specifically, the optimiser converges to a test unit allocation of $n_1= 71, n_2=24, n_3=5$, but were we to apply the conventional 4:2:1 rule, the test unit allocation would be: $n_1=57, n_2 = 29, n_3 = 14$. Whilst our results are reasonably consistent with broad guidelines provided in the literature, our framework finds a solution tailored to the specific quadratic relationship being analysed.


\subsection{\textit{Power Law Life-Stress Relationship}}
\label{sec:Power}

\begin{figure}[!h]
\centering
\includegraphics[width=11cm, height=8cm]{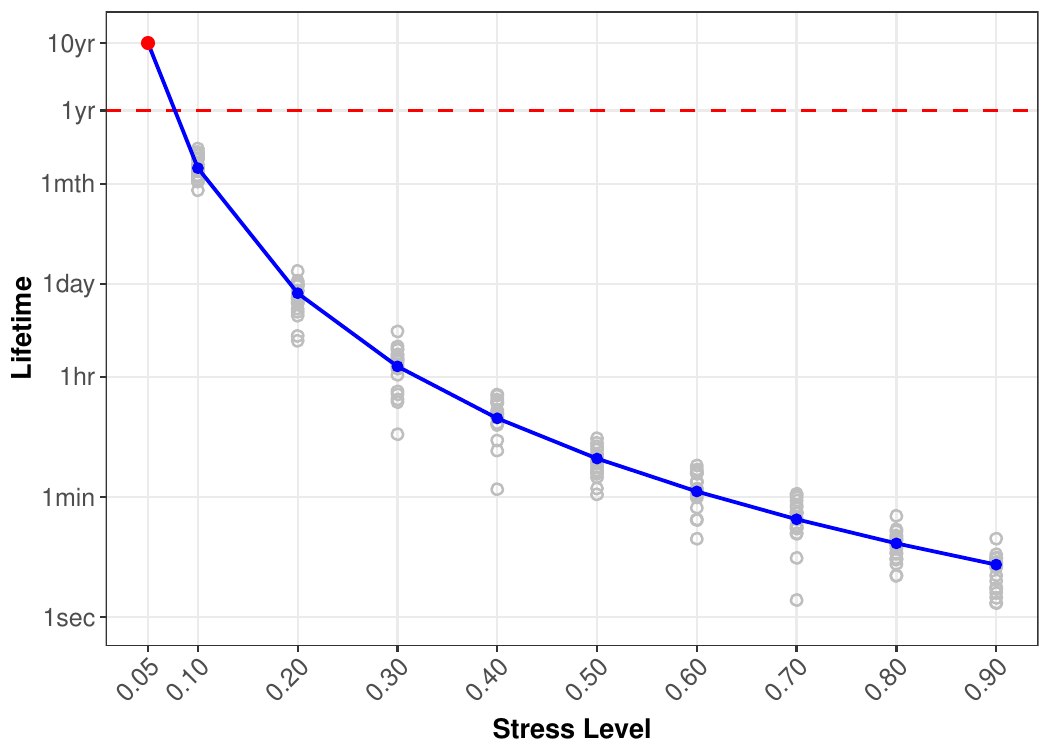}
\caption{Sample preliminary accelerated test data depicting a power law trend. All lifetimes are uncensored.}
\label{Fig:data_power}
\end{figure}

Next, we discuss a power law life-stress relationship. Figure \ref{Fig:data_power} depicts preliminary data with a power law trend, so we fit a model of the form: $\mu(S) = \beta_0 + \beta_1\log(S) $, giving estimated coefficients of $\hat{\beta_0} = -6.9, \hat{\beta_1} = -6.2$ and $ \hat{\sigma} = 0.5$.

\begin{table}[!h]
\small
\centering
\caption{Optimal test plan configurations for power law AFT model.}
\label{Tab:results_power}
\begin{tabular}{c|cccccc|c|c|c|c|c} 

\multirow{2}{*}{$N$} & \multicolumn{6}{c|}{Stress Levels}   &\multicolumn{1}{c|}{Min.} &\multicolumn{1}{c|}{Mean} &  \multicolumn{1}{c|}{Std.} & \multirow{2}{*}{$p_{0.05}$} & \multirow{2}{*}{$p_{0.95}$} \\
 & $S_1$ & $S_2$ & $S_3$ & $S_4$ & $S_5$ & $S_6$ & \multicolumn{1}{c|}{RMSE} & \multicolumn{1}{c|}{RMSE} & \multicolumn{1}{c|}{Error} &&  \\
\hline
&&&&&&&&&&\\[-0.45cm]
\multirow{2}{*}{2} & 0.10 & 0.89 & \multirow{2}{*}{---} & \multirow{2}{*}{---} & \multirow{2}{*}{---} & \multirow{2}{*}{---} & \multirow{2}{*}{\textbf{7,239}}  & \multirow{2}{*}{7,945}& \multirow{2}{*}{191}& \multirow{2}{*}{7,641} & \multirow{2}{*}{8,274}\\
&(77)&(23)&&&&&&&&&\\
&&&&&&&&&&&\\[-0.45cm]
\hline
&&&&&&&&&&&\\[-0.45cm]
\multirow{2}{*}{3} & 0.10 & 0.56 & 0.90 & \multirow{2}{*}{---} & \multirow{2}{*}{---}& \multirow{2}{*}{---}  & \multirow{2}{*}{7,331}  & \multirow{2}{*}{7,953}& \multirow{2}{*}{182} & \multirow{2}{*}{7,656} & \multirow{2}{*}{8,251}\\
&(77) & (01) & (22) &&&&&&&&\\
&&&&&&&&&&&\\[-0.45cm]
\hline
&&&&&&&&&&&\\[-0.45cm]
\multirow{2}{*}{4} & 0.10 & 0.81 & 0.90 & 0.90 & \multirow{2}{*}{---} & \multirow{2}{*}{---}  & \multirow{2}{*} {7,303}  & \multirow{2}{*}{7,902} & \multirow{2}{*}{184} & \multirow{2}{*}{7,578} & \multirow{2}{*}{8,198} \\
&(81) & (03) & (15) & (01) &&&&&&&\\
&&&&&&&&&&&\\[-0.45cm]
\hline
&&&&&&&&&&&\\[-0.45cm]
\multirow{2}{*}{5} & 0.10 & 0.86 & 0.90 & 0.90 & 0.90 & \multirow{2}{*}{---}  & \multirow{2}{*}{7,388}  &\multirow{2}{*}{\textbf{7,895}}& \multirow{2}{*}{174} & \multirow{2}{*}{7,619} & \multirow{2}{*}{8,183}\\
&(77) & (01) & (19) & (01) & (02) &&&&&&\\
&&&&&&&&&&&\\[-0.45cm]
\hline
&&&&&&&&&&&\\[-0.45cm]
\multirow{2}{*}{6} & 0.10 & 0.70 & 0.90 & 0.90 & 0.90 & 0.90  & \multirow{2}{*}{7,322} & \multirow{2}{*}{8,006}&\multirow{2}{*}{186} & \multirow{2}{*}{7,699} & \multirow{2}{*}{8,311}\\
&(80) & (02) & (07) & (01) & (02) & (08) &&&&& \\
&&&&&&&&&&&\\[-0.45cm]
\hline
&&&&&&&&&&&\\[-0.45cm]
\multirow{2}{*}{6\textsuperscript{*}} & 0.10 & 0.58 & 0.90  & 0.90  & 0.90  & \multirow{2}{*}{---}    & \multirow{2}{*}{7,295}  & \multirow{2}{*}{8,097}& \multirow{2}{*}{191} & \multirow{2}{*}{7,795} & \multirow{2}{*}{8,418}\\
&(80) & (01) & (02)   & (03)  & (14)   &    & & &&&\\
\end{tabular}
\end{table}

\begin{figure}[!h]
\centering
\includegraphics[width=11cm, height=8cm]{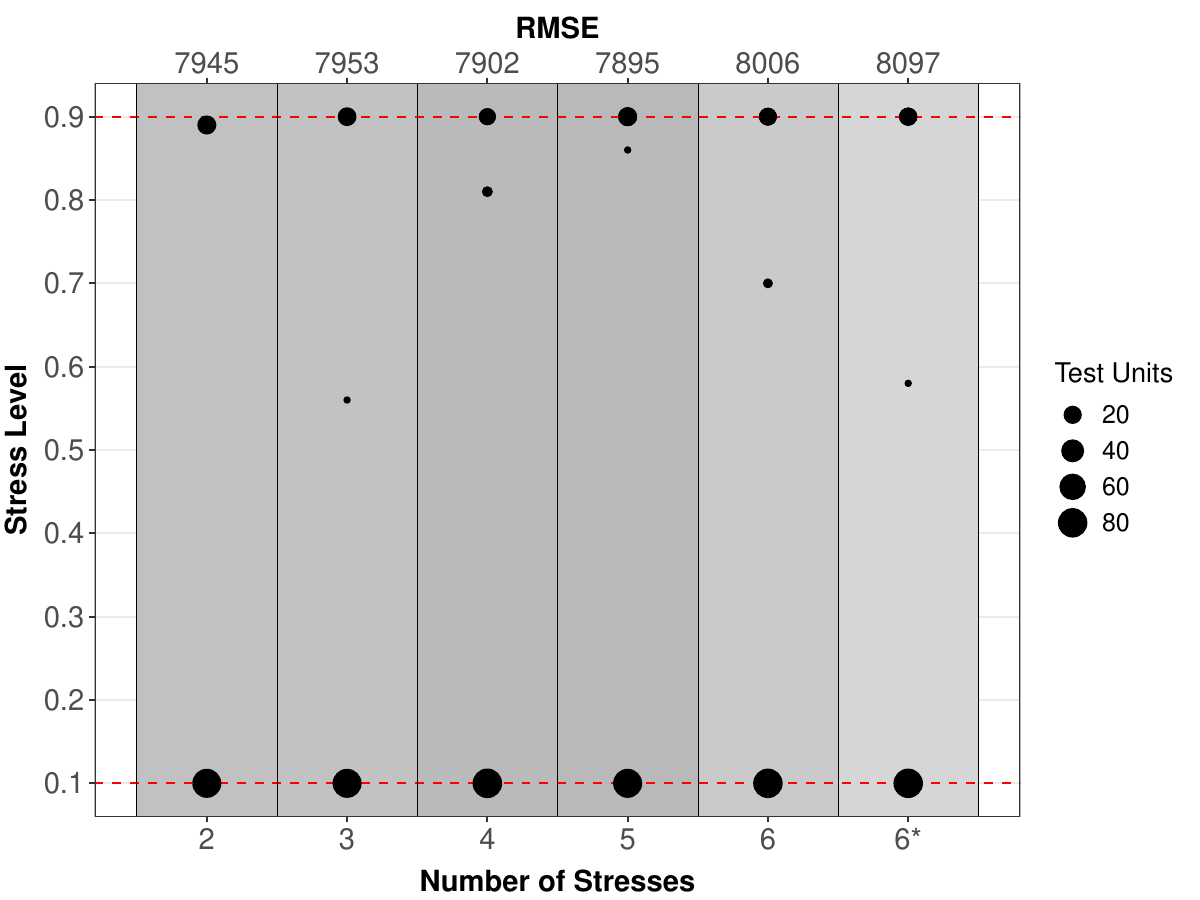}
\caption{Comparison of optimal test plans and RMSE for power law AFT model; greyscale colouring as per Figure \ref{Fig:Three_in_one_linear}.}
\label{Fig:Three_in_one_power}
\end{figure}

As shown in Table \ref{Tab:results_power} and Figure \ref{Fig:Three_in_one_power}, the results from the power law model analysis are very similar to the previous linear model analysis (see Section \ref{sec:Linear}), i.e., the optimiser converges to a 2-point design, where the majority of the test units are allocated to the lower stress. (\cite{Meeker:1975} discuss a Weibull power law example with similar results.) The similarity with the linear model is to be expected since the power law is also a two-parameter regression model that is linear on the log-$S$ scale as per Appendix \ref{App:A}. The main difference between the linear and power law analyses is the level of the optimal low stress. We saw from the linear model analysis, that the optimal low stress was $S_1=0.19$. As per Figure \ref{Fig:data_linear}, $95\%$ of the lifetimes at $S=0.1$ were right-censored, so the optimiser favoured a higher stress of $S=0.19$. However, with the power law lifetime data (Figure \ref{Fig:data_power}), all of the lifetimes are uncensored, and the optimiser has selected the lowest stress of $S=0.1$ as it is not impaired by censoring, while providing more information than $S = 0.19$ (in that it is further away from the high stress). This shows that the proportion of censored lifetimes (dictated by the experimental duration) influences the optimal low stress level of the test plan.  Appendices \ref{App:B_test_duration_100} and \ref{App:B_test_duration_200} show that, for the linear AFT model, an optimal low stress of $S_1=0.23$ is obtained when the experiment duration, $t_e$, is reduced from one year to 6 months. 

Our analysis of Weibull AFT models for linear, quadratic and power law functional forms has shown that our proposed simulation-based optimisation framework is sufficiently general to optimise a wide range of test plans. Moreover, where previous literature has provided analytic results/guidelines, our optimiser provides similar results, while being more targeted to the specific parameter values at hand. As discussed previously, the main advantages of our approach is that it offers: (i) capability to handle more complicated test plans where analytical methods are not feasible, and (ii) flexibility to be applied across a range of practical ALT experiments. To this end, in the next section we utilise a case study to demonstrate how the framework is easily applied to real-world accelerated test data.


\section{Case Study - Electronic Component Lifetime}
\label{sec:Case_Study}

\begin{figure}[!h]
\centering
\includegraphics[width=11cm, height=8cm]{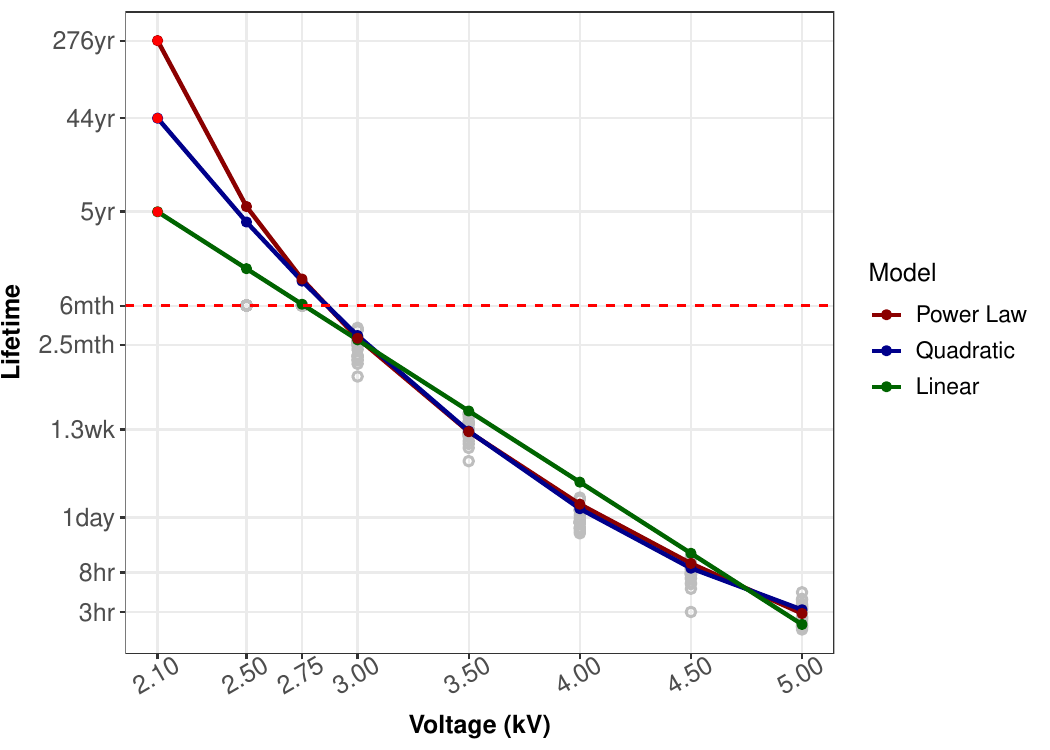}
\caption{Constant-stress accelerated test data with fitted Weibull reliability models.  All lifetimes at 2.5kV and 2.75kV are right censored. Estimated design-stress ($2.1$kV) lifetimes for each model are shown as a solid (red) point.}
\label{Fig:data_Case}
\end{figure}

We now demonstrate the practical application of our optimisation framework by utilising industrial ALT data from an industry partner of ours. In this study, the manufacturing company are interested in estimating the lifetimes of electronic components whose intended operating voltage is 2.1kV (kilovolt). An initial accelerated life test was previously conducted where $n=23$ components were tested at each of the following (constant) voltage stresses: $\{2.5, 2.75, 3, 3.5, 4, 4.5, 5\}$kV. All components were tested for a period of 6 months, or until failure, whichever came first. The lifetime data from this experiment, and a selection of fitted Weibull reliability models are given in Figure \ref{Fig:data_Case}.

Figure \ref{Fig:data_Case} shows that the linear, quadratic and power law models yield significantly different estimates at the design stress of 2.1kV (approximately 5 years, 44 years and 276 years respectively). The contrast in design-stress predictions highlights the importance of model selection within the simulation/optimisation framework. Of course, model selection is a fundamental aspect of many statistical analyses, but it is particularly important in the context of ALT, where model extrapolation is necessary. We can see that the quadratic and power law models appear to give reasonably similar estimates (albeit on the log-scale) at the accelerated stresses, but quite different estimates are obtained once the models are extrapolated to the design stress. 

To mitigate such estimation uncertainty, we have used a weighted Akaike information criterion (AIC) approach in this case study \citep[Chapter 2]{Burnham:2002}. We found that almost all ($99.97\%$) of the AIC weight was allocated to the quadratic model (indicating that the quadratic model provides a significantly better fit compared to the other models). Thus, for this particular analysis, we fit a quadratic model, and use the estimated model parameters:  $\hat{\beta_0} = 30.3, \hat{\beta_1} = -10.1, \hat{\beta_2} = 0.9, \hat{\sigma} = 0.2$ within our optimiser. In general, AIC weighted model-averaging is an effective way to account for model selection uncertainty, which could even be propagated into the optimisation algorithm, but that is beyond the scope of this paper. 

\begin{table}[!h]
\small
\centering
\caption{Optimal test plan configurations for quadratic AFT model (industrial data).}
\label{Tab:results_case_study}
\begin{tabular}{c|cccccc|c|c|c|c|c} 

\multirow{2}{*}{$N$} & \multicolumn{6}{c|}{Stress Levels}   &\multicolumn{1}{c|}{Min.} &\multicolumn{1}{c|}{Mean} &  \multicolumn{1}{c|}{Std.} & \multirow{2}{*}{$p_{0.05}$} & \multirow{2}{*}{$p_{0.95}$} \\
 & $S_1$ & $S_2$ & $S_3$ & $S_4$ & $S_5$ & $S_6$ & \multicolumn{1}{c|}{RMSE} & \multicolumn{1}{c|}{RMSE} & \multicolumn{1}{c|}{Error} &&  \\
\hline
&&&&&&&&&&\\[-0.45cm]
\multirow{2}{*}{3} & 2.91 & 4.01 & 5.00 & \multirow{2}{*}{---} & \multirow{2}{*}{---}& \multirow{2}{*}{---}  & \multirow{2}{*}{34,308} & \multirow{2}{*}{37,107}& \multirow{2}{*}{865} & \multirow{2}{*}{35,729} & \multirow{2}{*}{38,557}\\
&(76) & (63) & (22) &&&&&&&&\\
&&&&&&&&&&&\\[-0.45cm]
\hline
&&&&&&&&&&&\\[-0.45cm]
\multirow{2}{*}{4} & 2.89 & 3.95& 5.00 & 5.00 & \multirow{2}{*}{---} & \multirow{2}{*}{---}  & \multirow{2}{*} {34,798} & \multirow{2}{*}{\textbf{36,915}} & \multirow{2}{*}{901} & \multirow{2}{*}{35,478} & \multirow{2}{*}{38,496} \\
&(79) & (63) & (01) & (18) &&&&&&&\\
&&&&&&&&&&&\\[-0.45cm]
\hline
&&&&&&&&&&&\\[-0.45cm]
\multirow{2}{*}{5} & 2.90 & 3.91 & 5.00 & 5.00 & 5.00 & \multirow{2}{*}{---}  & \multirow{2}{*}{\textbf{33,715}}  &\multirow{2}{*}{37,284}& \multirow{2}{*}{887} & \multirow{2}{*}{35,830} & \multirow{2}{*}{38,765}\\
&(76) & (72) & (06) & (04) & (03) &&&&&&\\
&&&&&&&&&&&\\[-0.45cm]
\hline
&&&&&&&&&&&\\[-0.45cm]
\multirow{2}{*}{6} & 2.92 & 3.51 & 3.97 & 5.00 & 5.00 & 5.00 & \multirow{2}{*}{35,039}  & \multirow{2}{*}{37,675}&\multirow{2}{*}{888} & \multirow{2}{*}{36,271} & \multirow{2}{*}{39,175}\\
&(79) & (02) & (64) & (12) & (03) & (01) &&&& \\
&&&&&&&&&&&\\[-0.45cm]
\hline
&&&&&&&&&&&\\[-0.45cm]
\multirow{2}{*}{6\textsuperscript{*}} & 2.91 & 3.88 & 4.07  & 5.00   & \multirow{2}{*}{---}  & \multirow{2}{*}{---}    & \multirow{2}{*}{34,932} & \multirow{2}{*}{37,326}& \multirow{2}{*}{897} & \multirow{2}{*}{35,802} & \multirow{2}{*}{38,764}\\
&(82) & (35) & (26)   & (18)  &   &    & & &&&\\
\end{tabular}
\end{table}

\begin{figure}[!h]
\centering
\includegraphics[width=11cm, height=8cm]{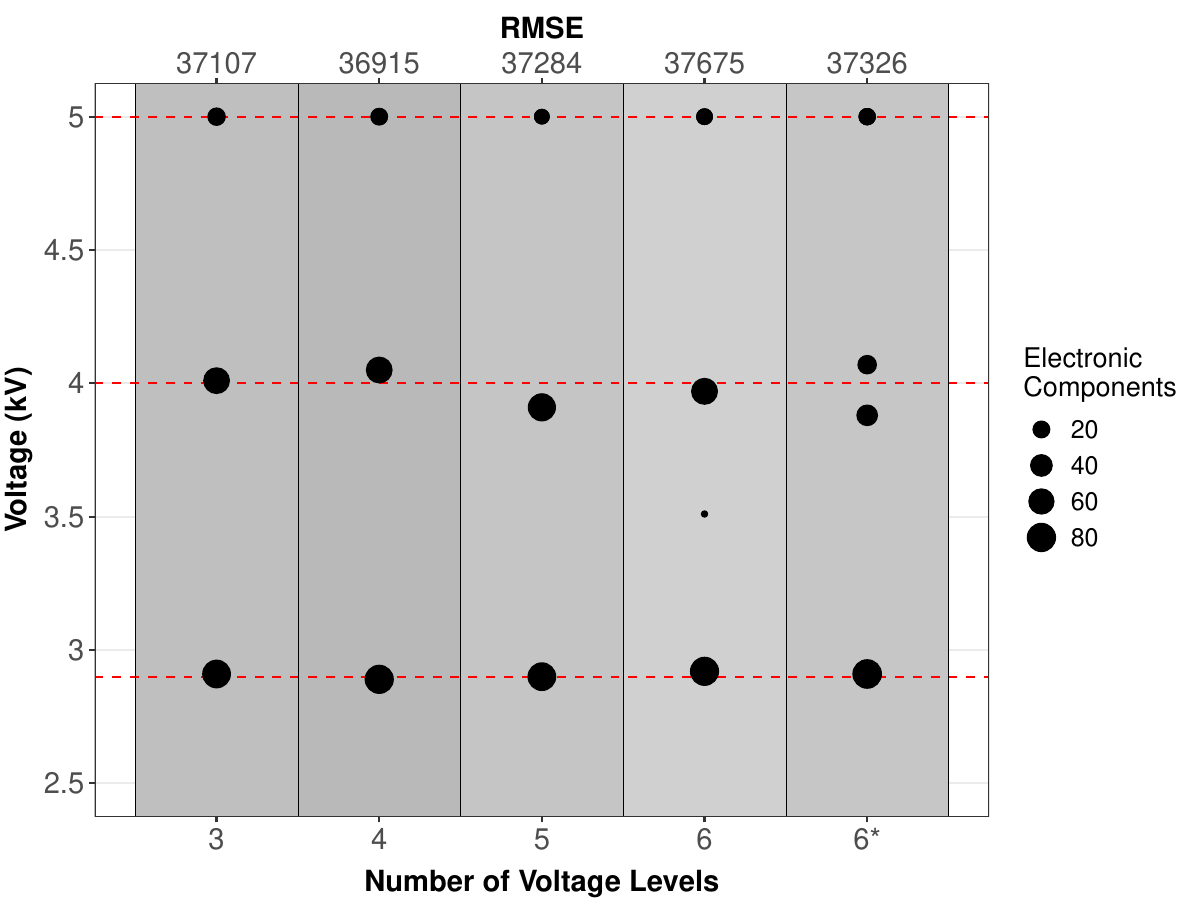}
\caption{Comparison of optimal test plans and RMSE for quadratic AFT model (industrial data); greyscale colouring as per Figure \ref{Fig:Three_in_one_linear}.}
\label{Fig:Three_in_one_case_study}
\end{figure}

\begin{figure}[!h]
\centering
\includegraphics[width=11cm, height=8cm]{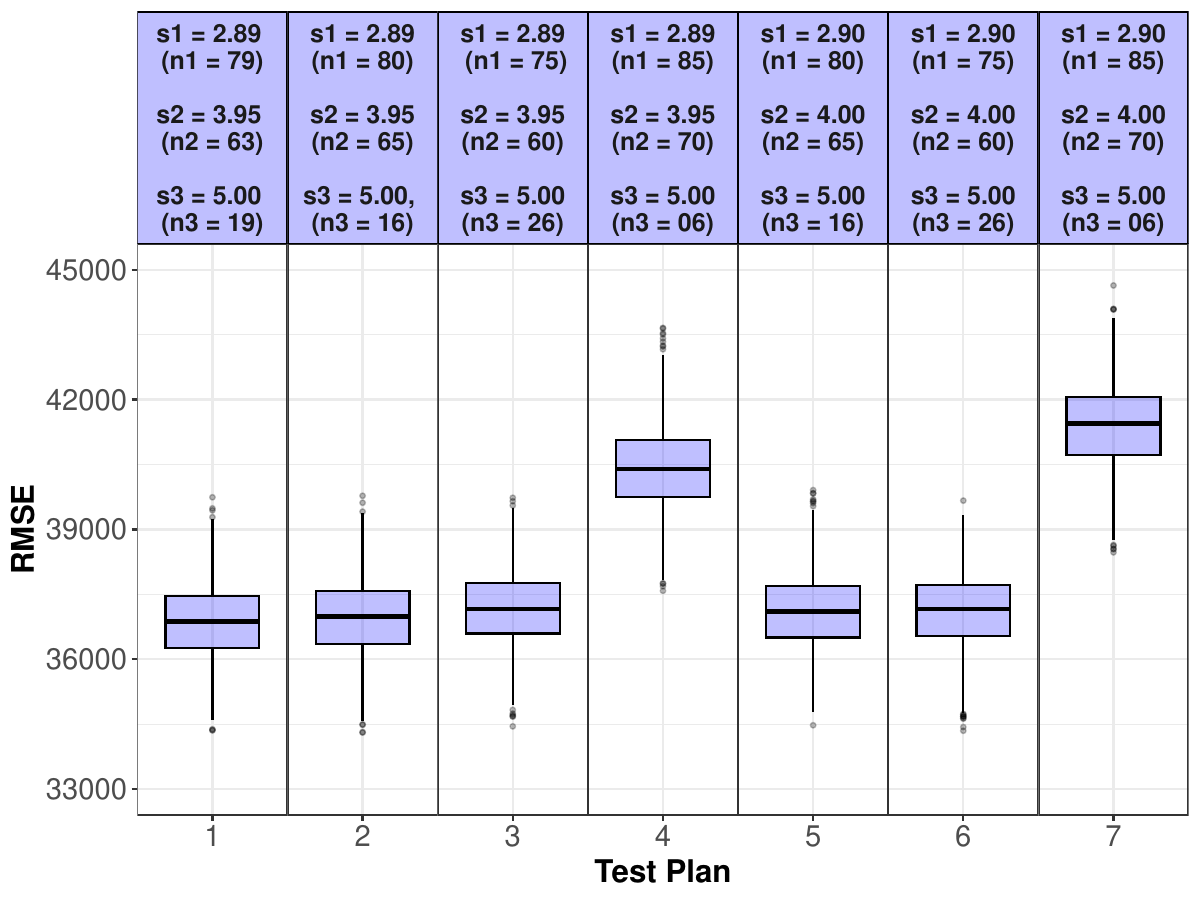}
\caption{Comparison of RMSE for 3-point quadratic test plans (industrial data).}
\label{Fig:compare_case_study}
\end{figure}

The optimisation results (Table \ref{Tab:results_case_study} and Figure \ref{Fig:Three_in_one_case_study}) show that the optimal test plan is a 3-point plan: $V_1\approx2.9$kV, $V_2 \approx 4$kV and $V_3=5$kV, with a corresponding test unit allocation of approximately 80:60:20. We can see from Figure \ref{Fig:compare_case_study} that, variations of 5 test units or so (test plans \#2 and \#3) and test unit variation coupled with modest stress rounding (test plans \#5 and \#6), provide comparable performance to the optimal test plan (test plan \#1). When the test units allocation is varied further (test plans \#4 and \#7), the out-of-sample prediction performance is reduced, i.e., the average RMSE is larger. However, in some applications, it could be that these higher-RMSE plans could still be viable, especially if the cost of running the experiment at the largest stress ($S_3$) is high. Indeed, the costs associated with the number of units, sizes of stresses, and experimental duration could be all be incorporated into a more general objective function than the purely predictive-performance-based RMSE objective function in (\ref{Eq:RMSE_sim}).

Overall, the findings from this application align broadly with our previous (simulated) quadratic model analysis (Section \ref{sec:Quad}) and the 4:2:1 rule from previous literature. Our Section \ref{sec:Quad} results assign 71\%, 24\%, and 5\% to each of a low, medium, and high stress, whereas the 4:2:1 rule implies allocations of 57\%, 29\%, and 14\% respectively. For the electronic component data in the current section, the allocation is 49\%, 39\%, and 12\%, respectively, to voltages $V_1 \approx 2.9$kV (low), $V_2 \approx 4$kV (medium), and $V_3 = 5$kV (high). Therefore, from a very generic perspective, the guideline is that more devices are allocated to lower stresses (as was also the case for the linear and power law relationships). While this general advice may be sufficient in some situations, the specific proportions and positions of the stresses will depend on the exact problem at hand. Our optimisation framework can be applied to configure any application-specific test plan, and such a plan can indeed be statistically superior than using only broad guidelines.


\section{Discussion}
\label{sec:Discuss}

Estimating the lifetime of a component at its intended operating stress (design stress) is a fundamental challenge in reliability engineering. The lifetime of many modern components is such that it can be quite difficult to observe component failure in normal operating conditions. Accelerated life testing (ALT) is a tried and tested method of obtaining lifetime information in a much shorter time by subjecting the components to higher-than-normal stress(es).  An essential aspect of ALT is the design of the accelerated testing experiment, i.e., the accelerated test plan. Specifically, reliability practitioners are interested in obtaining the test plan configuration that maximises the lifetime information from the experiment. In other words, a key goal in ALT is to seek optimal accelerated life test plans. In some cases, it is possible to optimise the test plan by analytical means, for example, simpler test plans with a limited number of stress levels. However, theoretical optimisation becomes very challenging when one is faced with a complicated design, with multiple inputs, perhaps coupled with a complex life-stress relationship.

In this work, we propose a simulation-based optimisation framework to optimise ALT test plans incorporating a single (constant) accelerating stress. Whilst optimisation of simpler test plans has a wide literature, discussion of complex test plans with a large number of stress levels is somewhat lacking. Utilising the differential evolution (DE) algorithm, we discuss an optimisation framework that offers a practical solution to optimising complex multipoint designs; this is particularly useful in settings where mathematical optimisation is intractable. We have proposed a general and flexible ALT modelling/optimisation framework that, irrespective of the complexity of the test plan, can be applied across a wide range of ALT testing scenarios. We have demonstrated this by applying our method to lifetime data from both simulation studies (linear, quadratic, and power law life-stress curves) and real industrial life testing data.

The optimisation framework is intended to aid ALT practitioners in designing and optimising accelerated test plans, and as such can be easily adapted to the user's specific testing requirements. In addition to specifying any life-stress relationship and lifetime distribution, the user can easily alter test plan parameters within the framework. For example, an analyst can modify the experiment duration, number of available test units, stress resolution, as well as incorporating experiment-specific constraints (for instance, test units may need to be assigned in specific batch/lot sizes). We consider the RMSE (root mean squared error) of out-of-sample model predictions (at a user-defined quantile) as our optimality criterion. However, the framework could be adapted to consider other optimality criteria of interest (such as the commonly-used D-optimality criterion). The simulation framework also allows one to assess a test plan's robustness to model and/or distribution misspecification. For example, if there is uncertainty regarding model selection, an analyst could generate lifetime data adhering to a specified (true) model, and then explore the performance of models that deviate from the true functional form. 

Our proposed ALT optimiser has been implemented using the statistical computing language \texttt{R}, and to accompany this paper, we provide open-source \texttt{R} code where users can utilise a range of adjustable test plan parameters to conduct ALT optimisations. In addition, the optimisation framework could be extended to incorporate additional business/experimental constraints.  For example, future work could include the business cost of running experiments for extended periods, and the inherent trade-off between experiment duration and lifetime censoring. To permit users to utilise and/or adapt the optimisation tool, our code is available at: \url{https://github.com/o-mcgrath/ALT_Optimisation}

Overall, our simulation-based accelerated test plan optimisation framework provides practical insights and test plan configurations tailored to the specific application at hand. Where analytic results/guidelines exist for simpler test plans in prior literature, our results are in line with these. However, our proposal has the advantage that it can optimise any test plan, irrespective of the life-stress relationship, lifetime distribution, censoring distribution, or real-world experimental constraints. This simple and effective tool is intended to supplement physical component testing and provide additional insight to the engineers and analysts working on real-life ALT problems. Incorporating this offline approach into one's analyses may lead to a reduction of expensive (in terms of both time and money) and destructive component testing, and ultimately contribute to improving component reliability.


\bibliography{References_McGrath_Burke}
\pagebreak
\section*{Appendix}
\appendix


\section{Empirical Models in Log-linear Form }
\label{App:A}

\begin{enumerate}[label=\textbf{\arabic*.}]
\item{} \textbf{Thermochemical Model}\\

$T(S) = A\exp\left(\dfrac{B}{T}-C S\right)$\\

Letting $\beta_0 = \log(A) + \dfrac{B}{T}$, $\beta_1 = -C$ gives: \fbox{$\log T(S) = \beta_0 + \beta_1S$}\\
\item{} \textbf{Arrhenius Model }\\

$T(S) = A\exp\left(\dfrac{B}{S}\right)$\\

Letting $\beta_0 = \log(A)$, $\beta_1 = B$ gives: \fbox{$\log T(S) = \beta_0 + \beta_1 \dfrac{1}{S}$}\\
\item{} \textbf{Power Law Model}\\

$T(S)= AS^{-N}$\\

Letting $\beta_0 = \log(A)$, $\beta_1 = -N$ gives: \fbox{$\log T(S) = \beta_0 + \beta_1\log(S)$}\\
\item{} \textbf{Exponential S\bm{$^{1/2}$} Model}\\

$T(S)= A\exp\left[ \dfrac{B - C \sqrt{S}}{T}\right]$\\

Letting $\beta_0 = \log(A) + \dfrac{B}{T}$, $\beta_1=\dfrac{C}{T}$ gives: \fbox{$\log T(S) = \beta_0 + \beta_1\sqrt{S}$}
\end{enumerate}
\newpage

\section{Further Optimisation Results (Linear AFT Model)}
\label{App:B}


\subsection{Increased Minimum Stress Distance - \texorpdfstring{$\bm{D=10^{-1}}$}{$D = 10^{-1}$}  }
\label{App:B_stress_distance}

\begin{table}[!h]
\small
\centering
\caption{Optimal test plans for linear AFT model. The minimum distance between consecutive stress levels is $D=10^{-1}$.}
\label{Tab:results_linear_stress_distance}
\begin{tabular}{c|cccccc|c|c|c|c|c} 

\multirow{2}{*}{$N$} & \multicolumn{6}{c|}{Stress Levels}   &\multicolumn{1}{c|}{Min.} &\multicolumn{1}{c|}{Mean} &  \multicolumn{1}{c|}{Std.} & \multirow{2}{*}{$p_{0.05}$} & \multirow{2}{*}{$p_{0.95}$} \\
 & $S_1$ & $S_2$ & $S_3$ & $S_4$ & $S_5$ & $S_6$ & \multicolumn{1}{c|}{RMSE} & \multicolumn{1}{c|}{RMSE} & \multicolumn{1}{c|}{Error} &&  \\
\hline
&&&&&&&&&&\\[-0.45cm]
\multirow{2}{*}{2} & 0.19 & 0.90 & \multirow{2}{*}{---} & \multirow{2}{*}{---} & \multirow{2}{*}{---} & \multirow{2}{*}{---} & \multirow{2}{*}{6,594}  & \multirow{2}{*}{\textbf{7,059}}& \multirow{2}{*}{169}& \multirow{2}{*}{6,794} & \multirow{2}{*}{7,337}\\
&(83)&(17)&&&&&&&&&\\
&&&&&&&&&&&\\[-0.45cm]
\hline
&&&&&&&&&&&\\[-0.45cm]
\multirow{2}{*}{3} & 0.19 & 0.45 & 0.90 & \multirow{2}{*}{---} & \multirow{2}{*}{---}& \multirow{2}{*}{---}  & \multirow{2}{*}{6,556} & \multirow{2}{*}{7,064}& \multirow{2}{*}{171} & \multirow{2}{*}{6,787} & \multirow{2}{*}{7,358}\\
&(85) & (01) & (14) &&&&&&&&\\
&&&&&&&&&&&\\[-0.45cm]
\hline
&&&&&&&&&&&\\[-0.45cm]
\multirow{2}{*}{4} & 0.20 & 0.70 & 0.80 & 0.90 & \multirow{2}{*}{---} & \multirow{2}{*}{---}  & \multirow{2}{*} {6,610}  & \multirow{2}{*}{7,095} & \multirow{2}{*}{164} & \multirow{2}{*}{6,809} & \multirow{2}{*}{7,362} \\
&(83) & (01) & (01) & (15) &&&&&&&\\
&&&&&&&&&&&\\[-0.45cm]
\hline
&&&&&&&&&&&\\[-0.45cm]
\multirow{2}{*}{5} & 0.20 & 0.41 & 0.70 & 0.80 & 0.90 & \multirow{2}{*}{---}  & \multirow{2}{*}{\textbf{6,511}} &\multirow{2}{*}{7,197}& \multirow{2}{*}{161} & \multirow{2}{*}{6,939} & \multirow{2}{*}{7,455}\\
&(77) & (02) & (01) & (01) & (19) &&&&&&\\
&&&&&&&&&&&\\[-0.45cm]
\hline
&&&&&&&&&&&\\[-0.45cm]
\multirow{2}{*}{6} & 0.20 & 0.31 & 0.60 & 0.70 & 0.80 & 0.90  & \multirow{2}{*}{6,651}  & \multirow{2}{*}{7,236}&\multirow{2}{*}{167} & \multirow{2}{*}{6,961} & \multirow{2}{*}{7,516}\\
&(77) & (05) & (01) & (02) & (01) & (14) &&&&& \\
&&&&&&&&&&&\\[-0.45cm]
\hline
&&&&&&&&&&&\\[-0.45cm]
\multirow{2}{*}{6\textsuperscript{*}} & 0.20 & 0.90 & \multirow{2}{*}{---}  & \multirow{2}{*}{---}   & \multirow{2}{*}{---}  & \multirow{2}{*}{---}    & \multirow{2}{*}{6,752}  & \multirow{2}{*}{7,179}& \multirow{2}{*}{167} & \multirow{2}{*}{6,917} & \multirow{2}{*}{7,463}\\
&(77) & (23) &   &  &   &    & & &&&\\
\end{tabular}
\end{table}


\subsection{Increased Test Units - \texorpdfstring{$\bm{n=200}$}{$n=200$} }
\label{App:B_test_units}

\begin{table}[!h]
\small
\centering
\caption{Optimal test plans for linear AFT model. The number of test units is $n= 200$.}
\label{Tab:results_linear_test_units}
\begin{tabular}{c|cccccc|c|c|c|c|c} 

\multirow{2}{*}{$N$} & \multicolumn{6}{c|}{Stress Levels}   &\multicolumn{1}{c|}{Min.} &\multicolumn{1}{c|}{Mean} &  \multicolumn{1}{c|}{Std.} & \multirow{2}{*}{$p_{0.05}$} & \multirow{2}{*}{$p_{0.95}$} \\
 & $S_1$ & $S_2$ & $S_3$ & $S_4$ & $S_5$ & $S_6$ & \multicolumn{1}{c|}{RMSE} & \multicolumn{1}{c|}{RMSE} & \multicolumn{1}{c|}{Error} &&  \\
\hline
&&&&&&&&&&\\[-0.45cm]
\multirow{2}{*}{2} & 0.20 & 0.90 & \multirow{2}{*}{---} & \multirow{2}{*}{---} & \multirow{2}{*}{---} & \multirow{2}{*}{---} & \multirow{2}{*}{4,556}  & \multirow{2}{*}{4,950}& \multirow{2}{*}{117}& \multirow{2}{*}{4,750} & \multirow{2}{*}{5,138}\\
&(175)&(25)&&&&&&&&&\\
&&&&&&&&&&&\\[-0.45cm]
\hline
&&&&&&&&&&&\\[-0.45cm]
\multirow{2}{*}{3} & 0.19 & 0.52 & 0.90 & \multirow{2}{*}{---} & \multirow{2}{*}{---}& \multirow{2}{*}{---}  & \multirow{2}{*}{4,640}  & \multirow{2}{*}{\textbf{4,841}}& \multirow{2}{*}{108} & \multirow{2}{*}{4,763} & \multirow{2}{*}{5,118}\\
&(172) & (01) & (27) &&&&&&&&\\
&&&&&&&&&&&\\[-0.45cm]
\hline
&&&&&&&&&&&\\[-0.45cm]
\multirow{2}{*}{4} & 0.20 & 0.90 & 0.90 & 0.90 & \multirow{2}{*}{---} & \multirow{2}{*}{---}  & \multirow{2}{*} {4,576}  & \multirow{2}{*}{4,953} & \multirow{2}{*}{117} & \multirow{2}{*}{4,767} & \multirow{2}{*}{5,140} \\
&(168) & (05) & (25) & (02) &&&&&&&\\
&&&&&&&&&&&\\[-0.45cm]
\hline
&&&&&&&&&&&\\[-0.45cm]
\multirow{2}{*}{5} & 0.20 & 0.88 & 0.90 & 0.90 & 0.90 & \multirow{2}{*}{---}  & \multirow{2}{*}{\textbf{4,546}}  &\multirow{2}{*}{4,946}& \multirow{2}{*}{109} & \multirow{2}{*}{4,764} & \multirow{2}{*}{5,116}\\
&(169) & (02) & (02) & (02) & (25) &&&&&&\\
&&&&&&&&&&&\\[-0.45cm]
\hline
&&&&&&&&&&&\\[-0.45cm]
\multirow{2}{*}{6} & 0.20 & 0.90 & 0.90 & 0.90 & 0.90 & 0.90  & \multirow{2}{*}{4,596}  & \multirow{2}{*}{5,015}&\multirow{2}{*}{115} & \multirow{2}{*}{4,827} & \multirow{2}{*}{5,205}\\
&(181) & (02) & (02) & (01) & (13) & (01) &&&&& \\
&&&&&&&&&&&\\[-0.45cm]
\hline
&&&&&&&&&&&\\[-0.45cm]
\multirow{2}{*}{6\textsuperscript{*}} & 0.20 & 0.88 & 0.90  & 0.90   & \multirow{2}{*}{---}  & \multirow{2}{*}{---}    & \multirow{2}{*}{4,567}  & \multirow{2}{*}{4,944}& \multirow{2}{*}{114} & \multirow{2}{*}{4,750} & \multirow{2}{*}{5,134}\\
&(172) & (01) &  (03) & (24)  &   &    & & &&&\\
\end{tabular}
\end{table}
\newpage


\subsection{Reduced Simulation Replicates - \texorpdfstring{$\bm{n_{\text{sim}} = 500}$}{$n_{\text{sim}} = 500$} }
\label{App:B_nsim}

\begin{table}[!h]
\small
\centering
\caption{Optimal test plans for linear AFT model. The number of simulation replicates is $n_{\text{sim}} = 500$.}
\label{Tab:results_linear_nsim}
\begin{tabular}{c|cccccc|c|c|c|c|c} 

\multirow{2}{*}{$N$} & \multicolumn{6}{c|}{Stress Levels}   &\multicolumn{1}{c|}{Min.} &\multicolumn{1}{c|}{Mean} &  \multicolumn{1}{c|}{Std.} & \multirow{2}{*}{$p_{0.05}$} & \multirow{2}{*}{$p_{0.95}$} \\
 & $S_1$ & $S_2$ & $S_3$ & $S_4$ & $S_5$ & $S_6$ & \multicolumn{1}{c|}{RMSE} & \multicolumn{1}{c|}{RMSE} & \multicolumn{1}{c|}{Error} &&  \\
\hline
&&&&&&&&&&\\[-0.45cm]
\multirow{2}{*}{2} & 0.20 & 0.90 & \multirow{2}{*}{---} & \multirow{2}{*}{---} & \multirow{2}{*}{---} & \multirow{2}{*}{---} & \multirow{2}{*}{6,315}& \multirow{2}{*}{\textbf{7,061}} & \multirow{2}{*}{235}& \multirow{2}{*}{6,680} & \multirow{2}{*}{7,464}\\
&(84)&(16)&&&&&&&&&\\
&&&&&&&&&&&\\[-0.45cm]
\hline
&&&&&&&&&&&\\[-0.45cm]
\multirow{2}{*}{3} & 0.19 & 0.85 & 0.90 & \multirow{2}{*}{---} & \multirow{2}{*}{---}& \multirow{2}{*}{---}  & \multirow{2}{*}{6,302} & \multirow{2}{*}{7,115}& \multirow{2}{*}{233} & \multirow{2}{*}{6,741} & \multirow{2}{*}{7,502}\\
&(84) & (06) & (10) &&&&&&&&\\
&&&&&&&&&&&\\[-0.45cm]
\hline
&&&&&&&&&&&\\[-0.45cm]
\multirow{2}{*}{4} & 0.20 & 0.89 & 0.90 & 0.90 & \multirow{2}{*}{---} & \multirow{2}{*}{---}  & \multirow{2}{*} {6,286} & \multirow{2}{*}{7,073} & \multirow{2}{*}{232} & \multirow{2}{*}{6,701} & \multirow{2}{*}{7,453} \\
&(88) & (07) & (04) & (01)) &&&&&&&\\
&&&&&&&&&&\\[-0.45cm]
\hline
&&&&&&&&&&&\\[-0.45cm]
\multirow{2}{*}{5} & 0.19 & 0.73 & 0.90& 0.90 & 0.90 & \multirow{2}{*}{---}  & \multirow{2}{*}{6,208}  &\multirow{2}{*}{7,086}& \multirow{2}{*}{246} & \multirow{2}{*}{6,699} & \multirow{2}{*}{7,507}\\
&(87) & (01) & (02) & (02) & (08) &&&&&&\\
&&&&&&&&&&&\\[-0.45cm]
\hline
&&&&&&&&&&&\\[-0.45cm]
\multirow{2}{*}{6} & 0.20 & 0.90 & 0.90 & 0.90 & 0.90 & 0.90  & \multirow{2}{*}{\textbf{6,150}}  & \multirow{2}{*}{7,095}&\multirow{2}{*}{235} & \multirow{2}{*}{6,703} & \multirow{2}{*}{7,482}\\
&(81) & (03) & (02) & (12) & (01) & (01) &&&&& \\
&&&&&&&&&&&\\[-0.45cm]
\hline
&&&&&&&&&&&\\[-0.45cm]
\multirow{2}{*}{6\textsuperscript{*}} & 0.20 & 0.84 & 0.90 & \multirow{2}{*}{---}   & \multirow{2}{*}{---}  & \multirow{2}{*}{---}    & \multirow{2}{*}{6,174} & \multirow{2}{*}{7,100}& \multirow{2}{*}{236} & \multirow{2}{*}{6,724} & \multirow{2}{*}{7,507}\\
&(83) & (01) & (16)   &  &   &    & & &&&\\
\end{tabular}
\end{table}


\subsection{Reduced Test Duration - \texorpdfstring{$\bm{t_{e} = 6}$ months}{$t_{e} = 6$ months} (100 test units) }
\label{App:B_test_duration_100}

\begin{table}[!h]
\small
\centering
\caption{Optimal test plans for linear AFT model. The test duration is six months, and the number of test units is $n=100$.}
\label{Tab:results_linear_test_duration_100}
\begin{tabular}{c|cccccc|c|c|c|c|c} 

\multirow{2}{*}{$N$} & \multicolumn{6}{c|}{Stress Levels}   &\multicolumn{1}{c|}{Min.} &\multicolumn{1}{c|}{Mean} &  \multicolumn{1}{c|}{Std.} & \multirow{2}{*}{$p_{0.05}$} & \multirow{2}{*}{$p_{0.95}$} \\
 & $S_1$ & $S_2$ & $S_3$ & $S_4$ & $S_5$ & $S_6$ & \multicolumn{1}{c|}{RMSE} & \multicolumn{1}{c|}{RMSE} & \multicolumn{1}{c|}{Error} &&  \\
\hline
&&&&&&&&&&\\[-0.45cm]
\multirow{2}{*}{2} & 0.22 & 0.90 & \multirow{2}{*}{---} & \multirow{2}{*}{---} & \multirow{2}{*}{---} & \multirow{2}{*}{---} & \multirow{2}{*}{7,046}  & \multirow{2}{*}{7,760}& \multirow{2}{*}{193}& \multirow{2}{*}{7,447} & \multirow{2}{*}{8,085}\\
&(80)&(20)&&&&&&&&&\\
&&&&&&&&&&&\\[-0.45cm]
\hline
&&&&&&&&&&&\\[-0.45cm]
\multirow{2}{*}{3} & 0.24 & 0.90 & 0.90 & \multirow{2}{*}{---} & \multirow{2}{*}{---}& \multirow{2}{*}{---}  & \multirow{2}{*}{7,062} & \multirow{2}{*}{7,643}& \multirow{2}{*}{177} & \multirow{2}{*}{7,360} & \multirow{2}{*}{7,936}\\
&(81) & (01) & (18) &&&&&&&&\\
&&&&&&&&&&&\\[-0.45cm]
\hline
&&&&&&&&&&&\\[-0.45cm]
\multirow{2}{*}{4} & 0.23 & 0.75 & 0.90 & 0.90 & \multirow{2}{*}{---} & \multirow{2}{*}{---}  & \multirow{2}{*} {\textbf{6,961}}  & \multirow{2}{*}{7,711} & \multirow{2}{*}{179} & \multirow{2}{*}{7,405} & \multirow{2}{*}{7,998} \\
&(86) & (01) & (11) & (02) &&&&&&&\\
&&&&&&&&&&&\\[-0.45cm]
\hline
&&&&&&&&&&&\\[-0.45cm]
\multirow{2}{*}{5} & 0.23 & 0.90 & 0.90 & 0.90 & 0.90 & \multirow{2}{*}{---}  & \multirow{2}{*}{7,096} &\multirow{2}{*}{\textbf{7,626}}& \multirow{2}{*}{176} & \multirow{2}{*}{7,328} & \multirow{2}{*}{7,901}\\
&(81) & (01) & (02) & (15) & (01) &&&&&&\\
&&&&&&&&&&&\\[-0.45cm]
\hline
&&&&&&&&&&&\\[-0.45cm]
\multirow{2}{*}{6} & 0.24 & 0.84 & 0.90 & 0.90 & 0.90 & 0.90  & \multirow{2}{*}{7,065} & \multirow{2}{*}{7,656}&\multirow{2}{*}{181} & \multirow{2}{*}{7,364} & \multirow{2}{*}{7,960}\\
&(82) & (01) & (01) & (01) & (01) & (14) &&&&& \\
&&&&&&&&&&&\\[-0.45cm]
\hline
&&&&&&&&&&&\\[-0.45cm]
\multirow{2}{*}{6\textsuperscript{*}} & 0.23 & 0.85 & 0.90  & 0.90   & 0.90  & 0.90    & \multirow{2}{*}{7,125}  & \multirow{2}{*}{7,717}& \multirow{2}{*}{182} & \multirow{2}{*}{7,422} & \multirow{2}{*}{8,025}\\
&(76) & (01) & (01)   & (01)  & (04)   & (17)    & & &&&\\
\end{tabular}
\end{table}
\newpage

\subsection{Reduced Test Duration - \texorpdfstring{$\bm{t_{e} = 6}$ months}{$t_{e} = 6$ months} (200 test units) }
\label{App:B_test_duration_200}

\begin{table}[!h]
\small
\centering
\caption{Optimal test plans for linear AFT model. The test duration is six months, and the number of test units is $n=200$.}
\label{Tab:results_linear_test_duration_200}
\begin{tabular}{c|cccccc|c|c|c|c|c} 

\multirow{2}{*}{$N$} & \multicolumn{6}{c|}{Stress Levels}   &\multicolumn{1}{c|}{Min.} &\multicolumn{1}{c|}{Mean} &  \multicolumn{1}{c|}{Std.} & \multirow{2}{*}{$p_{0.05}$} & \multirow{2}{*}{$p_{0.95}$} \\
 & $S_1$ & $S_2$ & $S_3$ & $S_4$ & $S_5$ & $S_6$ & \multicolumn{1}{c|}{RMSE} & \multicolumn{1}{c|}{RMSE} & \multicolumn{1}{c|}{Error} &&  \\
\hline
&&&&&&&&&&\\[-0.45cm]
\multirow{2}{*}{2} & 0.23 & 0.90 & \multirow{2}{*}{---} & \multirow{2}{*}{---} & \multirow{2}{*}{---} & \multirow{2}{*}{---} & \multirow{2}{*}{4,948}  & \multirow{2}{*}{\textbf{5,286}}& \multirow{2}{*}{127}& \multirow{2}{*}{5,080} & \multirow{2}{*}{5,497}\\
&(169)&(31)&&&&&&&&&\\
&&&&&&&&&&&\\[-0.45cm]
\hline
&&&&&&&&&&&\\[-0.45cm]
\multirow{2}{*}{3} & 0.24 & 0.90 & 0.90 & \multirow{2}{*}{---} & \multirow{2}{*}{---}& \multirow{2}{*}{---}  & \multirow{2}{*}{4,976} & \multirow{2}{*}{5,372}& \multirow{2}{*}{123} & \multirow{2}{*}{5,172} & \multirow{2}{*}{5,579}\\
&(155) & (43) & (02) &&&&&&&&\\
&&&&&&&&&&&\\[-0.45cm]
\hline
&&&&&&&&&&&\\[-0.45cm]
\multirow{2}{*}{4} & 0.23 & 0.66 & 0.90 & 0.90 & \multirow{2}{*}{---} & \multirow{2}{*}{---}  & \multirow{2}{*} {\textbf{4,919}} & \multirow{2}{*}{5,308} & \multirow{2}{*}{128} & \multirow{2}{*}{5,104} & \multirow{2}{*}{5,522} \\
&(169) & (01) & (24) & (06) &&&&&&&\\
&&&&&&&&&&&\\[-0.45cm]
\hline
&&&&&&&&&&\\[-0.45cm]
\multirow{2}{*}{5} & 0.24 & 0.90 & 0.90 & 0.90 & 0.90 & \multirow{2}{*}{---}  & \multirow{2}{*}{4,985} &\multirow{2}{*}{5,392}& \multirow{2}{*}{119} & \multirow{2}{*}{5,198} & \multirow{2}{*}{5,585}\\
&(164) & (01) & (28) & (03) & (04) &&&&&&\\
&&&&&&&&&&&\\[-0.45cm]
\hline
&&&&&&&&&&&\\[-0.45cm]
\multirow{2}{*}{6} & 0.24 & 0.43 & 0.84 & 0.87 & 0.90 & 0.90  & \multirow{2}{*}{4,948} & \multirow{2}{*}{5,407}&\multirow{2}{*}{124} & \multirow{2}{*}{5,209} & \multirow{2}{*}{5,614}\\
&(167) & (01) & (01) & (01) & (21) & (09) &&&&& \\
&&&&&&&&&&&\\[-0.45cm]
\hline
&&&&&&&&&&&\\[-0.45cm]
\multirow{2}{*}{6\textsuperscript{*}} & 0.23 & 0.56 & 0.90  & \multirow{2}{*}{---}   & \multirow{2}{*}{---}  & \multirow{2}{*}{---}    & \multirow{2}{*}{4,942} & \multirow{2}{*}{5,482}& \multirow{2}{*}{126} & \multirow{2}{*}{5,265} & \multirow{2}{*}{5,683}\\
&(159) & (09) & (32)   &  &   &    & & &&&\\
\end{tabular}
\end{table}

\end{document}